\documentclass[english,a4paper]{article}
\usepackage{dcolumn}
\usepackage{bm}
\usepackage{graphicx}
\usepackage{amsmath}
\usepackage{amsfonts}
\usepackage{amssymb}
\usepackage{amsthm}
\usepackage{amstext}
\usepackage{amsbsy}
\usepackage{amsopn}
\usepackage{amscd}
\usepackage{amsxtra}

\usepackage{dsfont} 

\def\C{\mathbb C}
\def\R{\mathbb R}

\def\N{\mathbb N}

\def\phi{\varphi}
\def\epsilon{\varepsilon}

\def\sn{\textrm sn}

\newcommand{\bra}[1]{\langle #1 |}
\newcommand{\ket}[1]{| #1 \rangle }
\newcommand{\braket}[2]{\langle #1| #2 \rangle }

\newcommand{\Pauli}[1]{\hat{\sigma}_{#1}}   
\newcommand{\vPauli}{\hat{\vec\sigma}}      

\newcommand{\hH}{\hat{H}}                   
\newcommand{\hU}{\hat{U}}                   
\newcommand{\hV}{\hat{V}}                   

\newcommand{\1}{\mathds{1}}                 
\newcommand{\sgn}{\mathrm{sgn}}             

\newcommand{\cn}{\mathrm{cn}}               
\newcommand{\dn}{\mathrm{dn}}               

\newcommand{\nd}{\mathrm{nd}}               

\everymath{\displaystyle}

\begin{document}

\title{Application of the Pontryagin Maximum Principle to the robust time-optimal control of two-level quantum systems}
\author{O. Fresse-Colson\footnote{Laboratoire Interdisciplinaire Carnot de
Bourgogne (ICB), UMR 6303 CNRS-Universit\'e Bourgogne Europe, 9 Av. A.
Savary, BP 47 870, F-21078 Dijon Cedex, France}, S. Gu\'erin\footnote{Laboratoire Interdisciplinaire Carnot de
Bourgogne (ICB), UMR 6303 CNRS-Universit\'e Bourgogne Europe, 9 Av. A.
Savary, BP 47 870, F-21078 Dijon Cedex, France}, Xi Chen\footnote{Instituto de Ciencia de Materiales de Madrid (CSIC), Cantoblanco, E-28049 Madrid, Spain}, D. Sugny\footnote{Laboratoire Interdisciplinaire Carnot de Bourgogne (ICB), UMR 6303 CNRS-Universit\'e Bourgogne Europe, 9 Av. A.
Savary, BP 47 870, F-21078 Dijon Cedex, France, dominique.sugny@u-bourgogne.fr}}

\maketitle

\begin{abstract}
We study the time-optimal robust control of a two-level quantum system subjected to field inhomogeneities. We apply the Pontryagin Maximum Principle and we introduce a reduced space onto which the optimal dynamics is projected down. This reduction leads to a complete analytical derivation of the optimal solution in terms of elliptic functions and elliptic integrals. Necessary optimality conditions are then obtained for the original system. These conditions are verified numerically and lead to the optimal control protocol. Various examples, ranging from state-to-state transfer to the generation of a Not gate, illustrate this study. The connection with other geometric optimization approaches that have been used to solve this problem is also discussed.
\end{abstract}

\section{Introduction}
Recently, remarkable progress has been made in the development of Quantum Optimal Control Theory (QOCT), which aims to manipulate quantum systems by external electromagnetic fields in the best possible way~\cite{glaserreview,brifreview,roadmap,alessandro}. Today, QOCT has become a key tool in quantum technologies with applications ranging from atomic~\cite{borzi,frank,dupont2021} and molecular physics~\cite{brifreview,RMP} to superconducting circuits, Nuclear Magnetic Resonance~\cite{glaserreview,kuprov}  and NV centers~\cite{rembold}. Various methods have been developed to design optimal control procedures. Analytical solutions can be found for low-dimensional systems~\cite{tutorial}, while numerical optimization algorithms are used in the high-dimensional case~\cite{grape,reichkrotov,gross,bryson}. These approaches are directly or indirectly related to the Pontryagin Maximum Principle (PMP)~\cite{pont,liberzon} which provides a rigorous mathematical framework for QOCT~\cite{boscain21}. The PMP has recently been applied to a number of fundamental quantum control problems, both for closed~\cite{boscain21,boscain,hegerfeldt:2013,khanejaspin,khaneja:2002,garon,dionis2023,stefanatos2024,stefanatos2023} and open quantum systems~\cite{lapert:2013,lapertprl,stefanatos2009,stefanatos2005,lewalle2023,lin2020} of low dimension. Finally, we point out that other approaches can be used to find open-loop control processes such as adiabatic~\cite{adiabaticreview} and shortcut to adiabaticity methods~\cite{guery2019,stefanatos2020} or composite pulses~\cite{compvitanov}.

Despite these impressive results, there are still a number of issues that need to be addressed by QOCT~\cite{roadmap}, in particular to bridge the gap between theory and experiment. A major limitation of most studies is that optimal control is designed in an open-loop framework without any feedback from the experiment. Incorporating robustness conditions into the optimal control design is a solution to ensure the experimental effectiveness of the process. Such robustness conditions to experimental inaccuracies can be considered explicitly in the static case where the value of one or more Hamiltonian parameters is not exactly known, but is assumed to be constant in time~\cite{khalid2023,koswara2014,koswara2021,kosut2013,gauguet,schirmer2025}. By studying the simultaneous control of an ensemble of quantum systems that differ from the value of this parameter, a robust process can be found~\cite{kobzar2008,kobzar:2012,nimbalkar:2012}. Another option is to use perturbation theory, such as Dyson series, to express the derivatives of the fidelity with respect to this parameter~\cite{muga2012,daemsprl}. This derivation has the advantage of limiting the complexity of the system to be controlled, allowing the use of analytical or geometric methods. This idea has been followed in a number of recent papers for the robust optimal control of two-level control systems~\cite{vandamme:2017,dridi2020,laforgue2022,dridi2024,meri2023,zeng2018,zeng2019,STAnjp,vandamme}. Some solutions have been obtained either from a geometric description of the control~\cite{zeng2018,zeng2019} or using a Lagrangian formulation of the optimization problem called RIO, for Robust Inversion Optimization~\cite{dridi2020,laforgue2022,dridi2024}. Such derivations give a physically insightful description of the optimal process, but they explicitly assume control properties that have not been verified in a generic way. In this paper, we propose to revisit this problem by applying the PMP to the robust time-optimal control of a two-level quantum system. In order to analytically express the optimal solution, we define a reduced space onto which the original optimal dynamics is projected down. We obtain necessary optimality conditions, which can be verified in a second step for the initial system. We show how this method can be used for state-to-state transfer and for the generation of a Not gate. Finally, we describe the connection between this method, RIO and the geometric optimization approach.

The remainder of this paper is organized as follows. The
control problem is presented in Sec.~\ref{sec2}, together with a derivation of the robustness conditions. In Sec.~\ref{sec3}, we show how to apply the PMP in this case and we define a reduced dynamical space. Regular and singular trajectories are described in Sec.~\ref{sec4new}. Section~\ref{sec4} is dedicated to the integration of the dynamics in this reduced space in terms of elliptic functions and integrals. This method is used in Sec.~\ref{sec5} and~\ref{sec6} for state-to-state transfer and Not gate generation, respectively. Conclusion and prospective views are given in Sec.~\ref{sec8}. Additional analytical results are provided in the appendices.

\section{The model system}\label{sec2}
We consider a two-level quantum system whose dynamics are governed in a given rotating frame and within the Rotating Wave Approximation by the following Hamiltonian
\begin{equation}
    \hH = \hH_0 + \alpha\hV = \frac{1}{2} \begin{pmatrix}
        -\Delta(t) & (1 + \alpha) \Omega(t) \\
        (1 + \alpha) \Omega(t) & \Delta(t)
    \end{pmatrix},
\end{equation}
with $\hH_0 = [-\Delta(t) \Pauli{z} + \Omega(t) \Pauli{x}] / 2$ and $\hV = \Omega(t) \Pauli{x} / 2$. The two-dimensional Hilbert space of the system $\mathcal{H}=\mathbb{C}^2$ is spanned by the canonical basis $\{|0\rangle,|1\rangle\}$. This system has two control parameters: $\Delta(t)$, the detuning between the frequency of the external electromagnetic field and the frequency transition of the quantum system, and $\Omega(t)$, the Rabi frequency which is proportional to the field amplitude. The term $\hV$, which describes field inhomogeneities, is a perturbation of $\hH_0$ with $|\alpha|\ll 1$ a small real parameter. We assume that the two control parameters satisfy the constraints $|\Delta(t)|\leq \Delta_0$ and $|\Omega(t)|\leq \Omega_0$ where the two bounds $\Delta_0$ and $\Omega_0$ are fixed by the experimental setup. The goal is to control this system in minimum time while being robust to field inhomogeneities, i.e. with respect to the parameter $\alpha$. In this study, we consider the definition of robustness used in~\cite{muga2012,daemsprl,dridi2020,laforgue2022} and briefly recalled below.

The state $\ket{\psi}$ of the system is the solution of the time-dependent Shr\"odinger equation, $i \ket{\dot{\psi}} = \hH \ket{\psi}$ (in units where $\hbar=1$). We consider its expansion in power series of the parameter $\alpha$ as
\begin{equation}
    \ket{\psi}= \sum_{n \in \N} \alpha^n \ket{\psi_n}= \ket{\psi_0} + \alpha \ket{\psi_1} + \alpha^2 \ket{\psi_2} + \dots
\end{equation}
The kets $\ket{\psi_n}$ satisfy the following dynamics
\begin{eqnarray*}
& & i\ket{\dot{\psi}_0}= \hH_0 \ket{\psi_0}, \\
& & i \ket{\dot{\psi}_n} = \hH_0 \ket{\psi_n} + \hV \ket{\psi_{n-1}},~\textrm{for}~n\geq 1.
\end{eqnarray*}
Note that the dynamics of $\ket{\psi_0}$ is only determined by $\hH_0$, and that the other kets $\ket{\psi_n}$ ($n\geq 1$) are not normalized. We use a Dyson series to find the expression of the different kets. At time $t_f$, we have
\begin{eqnarray*}
& & |\psi(t_f)\rangle = |\psi_0(t_f)\rangle -i \alpha\int_0^{t_f} dt \, \hU_0(t_f,t)\hV(t)|\psi_0(t)\rangle \\
& & -\alpha^2\int_0^{t_f}dt\int_0^tdt' \, \hU_0(t_f,t)\hV(t)\hU_0(t,t')\hV(t')|\psi_0(t')\rangle\\
& & +O(\alpha^3),
\end{eqnarray*}
where the evolution operator $\hat{U}_0$ is solution of $i\dot{\hat{U}}_0=\hat{H}_0\hat{U}_0$. Introducing the target state, $|\psi_f\rangle$, we arrive at
\begin{eqnarray*}
& & \langle\psi_f|\psi(t_f)\rangle = 1-i\alpha\int_0^{t_f}dt\,\langle \psi_0(t)|\hV(t)|\psi_0(t)\rangle  \\
& & -\alpha^2\int_0^{t_f}dt\int_0^tdt'\,\langle \psi_0(t)| \hV(t)\hU_0(t,t')\hV(t')|\psi_0(t')\rangle\\
& & +O(\alpha^3),
\end{eqnarray*}
where we use $\ket{\psi_f} = \ket{\psi_0(t_f)}$. We define the real function $e(t)$ as $e(t)=\langle \psi_0(t)|\hV(t)|\psi_0(t)\rangle$. The propagator $\hat{U}_0$ can be expressed as
$$
\hat{U}_0(t,t')=|\psi_0(t)\rangle\langle\psi_0(t')|+|\psi_\perp(t)\rangle\langle\psi_\perp(t')|,
$$
where $|\psi_\perp(t)\rangle$ is a state orthogonal to $|\psi_0(t)\rangle$, $\langle \psi_0(t)|\psi_\perp(t)\rangle =0$. We obtain
\begin{eqnarray*}
& & \langle\psi_f|\psi(t_f)\rangle = 1-i\alpha\int_0^{t_f}dt \, e(t)-\alpha^2\int_0^{t_f}dt \, e(t)\int_0^t dt' \, e(t') \\
& & -\alpha^2\int_0^{t_f} dt \, \bar{f}(t)\int_0^t dt' \, f(t')+O(\alpha^3),
\end{eqnarray*}
with $f(t)=\langle\psi_\perp(t)|\hV(t)|\psi_0(t)\rangle$. The latter equation can be simplified into
\begin{eqnarray*}
& & \langle\psi_f|\psi(t_f)\rangle = 1-i\alpha\int_0^{t_f} dt\, e(t) \\
& & - \frac{\alpha^2}{2}\left(\int_0^{t_f} dt \, e(t)\right)^2-\frac{\alpha^2}{2} \left|\int_0^{t_f}dt \, f(t)\right|^2+O(\alpha^3).
\end{eqnarray*}
For a state-to-state robust population transfer, the fidelity $\mathcal{F}_s$ of the control is defined as $\mathcal{F}_s = |\braket{\psi_f}{\psi(t_f)}|^2$. At order 2 in $\alpha$, the fidelity can be written as $\mathcal{F}_s=1-\alpha^2\left|\int_0^{t_f}dt \, f(t)\right|^2$ where we omit the term $O(\alpha^3)$. We deduce that the control is robust at order 2 in $\alpha$ if $\left|\int_0^{t_f}dt \, f(t)\right|=0$ i.e. if $$\Re\left[\int_0^{t_f} dt\, f(t)\right]=\Im\left[\int_0^{t_f}dt\, f(t)\right]=0.$$
For a qubit gate in $SU(2)$,  the fidelity $\mathcal{F}_g$ is given by $\mathcal{F}_g=\Re[\braket{\psi_f}{\psi(t_f)}]$ where the Cayley-Klein parameters are used to represent the evolution operator as a ket~\cite{laforgue2022}. We deduce that the implementation of the gate is robust if the conditions $E(t_f)=F(t_f)=0$ are both satisfied with $E(t)=\int_0^{t} dt' \, e(t')$ and $F(t)=\int_0^{t} dt'\, f(t')$. Only the second relation $F(t_f)=0$ is needed for state-to-state transfers.



\section{Application of the PMP}\label{sec3}
An optimal robust control is a control that is both robust under the conditions given in Sec.~\ref{sec2} and optimal for a given cost functional. In this paper, we study a time-optimal protocol for which the goal is to reach the target state in minimum time. The optimal control problem is formulated from the PMP. For details on the application of the PMP, we refer the interested reader to recent review papers~\cite{tutorial,boscain21}.

The PMP is applied in an extended space corresponding to the vectors of coordinates $(\ket{\psi_0},E,F)$ in order to account for robustness constraints. Starting from the state $(|\psi_0(0)\rangle,0,0)$, the goal is therefore to bring the system to $(|\psi_f\rangle,0,0)$ in minimum time. We denote by $(|\chi\rangle,p_e,p_f)$ the corresponding adjoint state with $p_e \in \R$ and $p_f = p_1 + i p_2 \in \C$. Note that $p_e=0$ for a state-to-state transfer. The PMP is then formulated from the pseudo-Hamiltonian $H_P$ as
$$
H_P=\Re\left(\bra{\chi}\dot{\psi}_0\rangle + p_e\dot{E}+p_f\dot{F}\right)-1,
$$
where only the normal extremals are considered~\cite{tutorial}. The final control time being free, we have $H_P=0$. The Pontryagin Hamiltonian $H_P$ can be transformed into
\begin{eqnarray*}
H_p&=&\Im[\bra{\chi}\hH_0\ket{\psi_0}] \\
& & + \frac{\Omega}{2}\big( p_e \bra{\psi_0}\Pauli{x}\ket{\psi_0}+\Re[p_f \bra{\psi_\perp}\Pauli{x}\ket{\psi_0}]\big)  - 1.
\end{eqnarray*}
The PMP states that the coordinates of the extremal state and of the corresponding adjoint state fulfill the Hamiltonian’s equations associated to the Hamiltonian $H_P$. We have
\begin{eqnarray*}
& & |\dot{\psi}_0\rangle =2 \partial_{\langle\chi |}H_P,~
\langle \dot{\chi}| = -2 \partial_{|\psi_0\rangle}H_P, \\
& & \dot{E}=\partial_{p_e}H_P,~\dot{p}_e=-\partial_E H_P,  \\
& & \dot{F}=2\partial_{p_f}H_P,~\dot{p}_f=-2\partial_F H_P,
\end{eqnarray*}
where the factor 2 is added for complex coordinates~\cite{tutorial}. For the state components, we deduce as expected that $|\psi_0\rangle$ is solution of the Schr\"odinger equation and that $\dot{E}=e$ and $\dot{F}=f$. For the dynamics of the adjoint state $|\chi\rangle$, we show in the Appendix~\ref{appder} that
\begin{equation}\label{eqadj}
    \langle\dot{\chi}| = i\bra{\chi}\hH_0 - \Omega \, p_e \bra{\psi_0}\Pauli{x} - \Omega \, p_f \bra{\psi_\perp}\Pauli{x}.
\end{equation}
Finally, since $H_P$ does not depend explicitly on $E$ and $F$, we obtain that $p_e$ and $p_f$ are constants of motion.

The optimal control problem can be solved by using the PMP formalism. In our case, the goal is to find the initial adjoint state $(|\chi(0)\rangle,p_e,p_f)$
such that the system described at time $t$ by $(|\psi(t)\rangle,E(t),F(t))$ reaches the target state $(|\psi_f\rangle,0,0)$ in minimum time. As such, the optimal solution cannot be integrated analytically. However, the complexity of the dynamics can be drastically reduced by considering the vectors $\vec{I}=(I_x,I_y,I_z)$ and $\vec{R}=(R_x,R_y,R_z)$ of $\mathbb{R}^3$ defined as
\begin{equation}\label{eqIR}
    \begin{aligned}
        \vec{I} &= \Im\bra{\chi}\vPauli\ket{\psi_0}, \\
        \vec{R} &= p_e \bra{\psi_0}\vPauli\ket{\psi_0}+\Re\left(p_f \bra{\psi_\perp}\vPauli\ket{\psi_0}\right).
    \end{aligned}
\end{equation}
In this reduced space, the dynamics are given by (see Appendix~\ref{appder} for details)
\begin{equation}\label{eqderIR}
    \begin{aligned}
        \dot{\vec{I}} &= \begin{pmatrix}
                0 & \Delta & 0 \\
                -\Delta & 0 & -\Omega \\
                0 & \Omega & 0
            \end{pmatrix}  \vec{I} + \Omega \begin{pmatrix}
                0 & 0 & 0 \\
                0 & 0 & -1 \\
                0 & 1 & 0
            \end{pmatrix} \vec{R}, \\
        \dot{\vec{R}} &= \begin{pmatrix}
                0 & \Delta & 0 \\
                -\Delta & 0 & -\Omega \\
                0 & \Omega & 0
            \end{pmatrix}  \vec{R}.
    \end{aligned}
\end{equation}
For a given set $(p_e,p_f)$, we observe that $R^2 = \vec{R}\cdot\vec{R}$ is a constant of motion. We can then use the projection of the system onto the $\vec{R}$-space, with a state moving on the sphere of radius $R$. The Pontryagin Hamiltonian can be rewritten as
\begin{equation}\label{hamdef}
    H_P = \frac{1}{2} \left(-\Delta I_z + \Omega I_x + \Omega R_x\right) - 1.
\end{equation}
Furthermore, the second-order robustness for both state-to-state transfer and quantum gate leads to a Necessary Condition to Robustness (NCR) in terms of $\vec{R}$ given by
\begin{equation}
    \label{eq:NCR}
    \mathcal{R}(t_f) = \int_0^{t_f} R_x(t) \, dt = 0.
\end{equation}
It is only a necessary condition since $E(t_f)=F(t_f)=0$ and $\Omega$ constant leads to $\mathcal{R}(t_f)=0$, but the reverse is not true.

We show in the following that the PMP can be solved exactly in the $\vec{R}$-space. The trajectories in this space are only a projection of the original optimal dynamics. Necessary conditions are thus obtained. In a second step, we check that some candidates for optimality given by the PMP with the $\vec{R}$-coordinates are also solutions of the control problem in the initial space.

\section{Regular and singular trajectories}\label{sec4new}
The PMP states that in the optimal case~\cite{boscain21,tutorial}, the problem can be restricted to the study of the Hamiltonian $H$ given by
$$
H = \max_{\substack{|\Delta(t)| \leq \Delta_0 \\ |\Omega(t)| \leq \Omega_0}} H_P.
$$
Here we are faced with a non-trivial optimal control problem in which the admissible control space is a rectangle~\cite{boscain2014}. We are interested in determining the times when controls are equal to their bounds $\pm \Delta_0$ and $\pm \Omega_0$ and when they may have values in the open intervals $]-\Delta_0,\Delta_0[$ and $]-\Omega_0,\Omega_0[$. Such controls correspond respectively to regular and singular solutions. More precisely, the maximization condition can be solved by introducing two switching functions $\Phi_\Delta$ and $\Phi_\Omega$ defined as
\begin{eqnarray*}
& & \Phi_\Delta = I_z, \\
& & \Phi_\Omega = I_x+R_x,
\end{eqnarray*}
and corresponding to the Hamiltonian terms in front of $\Delta$ and $\Omega$ in Eq.~\eqref{hamdef}. We deduce that
$\Delta(t)=\Delta_0\sgn[\Phi_\Delta]$ and $\Omega(t)=\Omega_0\sgn[\Phi_\Omega]$ when the corresponding switching function is different from zero. The associated control is of constant amplitude and is called a bang control. When the switching function vanishes at an isolated time $t_0$, the control changes sign and we get a bang-bang control sequence. The singular solution occurs when the switching function is zero on a time interval $[t_0,t_1]$. Note that $\Delta$ and $\Omega$ may be independently regular or singular. However, they cannot be simultaneously singular since in this case $H_P=-1$, which is in contradiction with the fact that $H_P=0$ when the final time is free.

We now derive the singular controls denoted $\Delta_s$ and $\Omega_s$ by exploiting the fact that the switching functions and their time derivatives are equal to zero. We have
$$
\begin{aligned}
\dot{\Phi}_\Delta &= \Omega (I_y+R_y), \\
\ddot{\Phi}_\Delta &= -\Delta\Omega (I_x+R_x)-\Omega^2 (I_z-2 R_z),
\end{aligned}
$$
and
$$
\begin{aligned}
\dot{\Phi}_\Omega &= \Delta (I_y+R_y), \\
\ddot{\Phi}_\Omega &= -\Delta^2(I_x+R_x)-\Delta\Omega(I_z+2R_z),
\end{aligned}
$$
where we assume that the other control is a constant because the singular-singular case is not optimal. For $\Delta_s$, starting from $\ddot{\Phi}_\Delta=0=-\Delta \Omega(I_x+R_x)-2\Omega^2 R_z$, we deduce from $H_P=0=\frac{\Omega}{2}(I_x+R_x)-1$ that
\begin{equation}\label{deltasing}
\Delta_s=-{\Omega_0}^2 R_z.
\end{equation}
In the case where $\Omega$ is singular, we obtain from $\ddot{\Phi}_\Omega=0=H_P$ that the only solution is $I_z=-\tfrac{2}{\Delta_0}$ and $R_z=\tfrac{1}{\Delta_0}$. This leads to $\Omega_s=0$. This control is not relevant in our study where the goal is to transfer the state to a target from the ground state of the system.

From the analysis of the Pontryagin Hamiltonian, we conclude that the optimal trajectory is the concatenation of bangs and singular arcs for $\Delta$ and bangs for $\Omega$. Such trajectories are respectively called regular and singular  in the rest of the paper.

\section{Integration of the optimal trajectories}\label{sec4}
In this section, we study the optimal dynamics in the $\vec{R}$-space. We show that the trajectories can be described by cosine and Jacobi functions in the regular and singular cases, respectively. The NCR can be expressed as a sum of elliptic integrals for the singular case, and of cosine functions for the regular case.

    \subsection{The singular case}
\begin{figure}[htbp]
\begin{center}
\includegraphics[width=8cm]{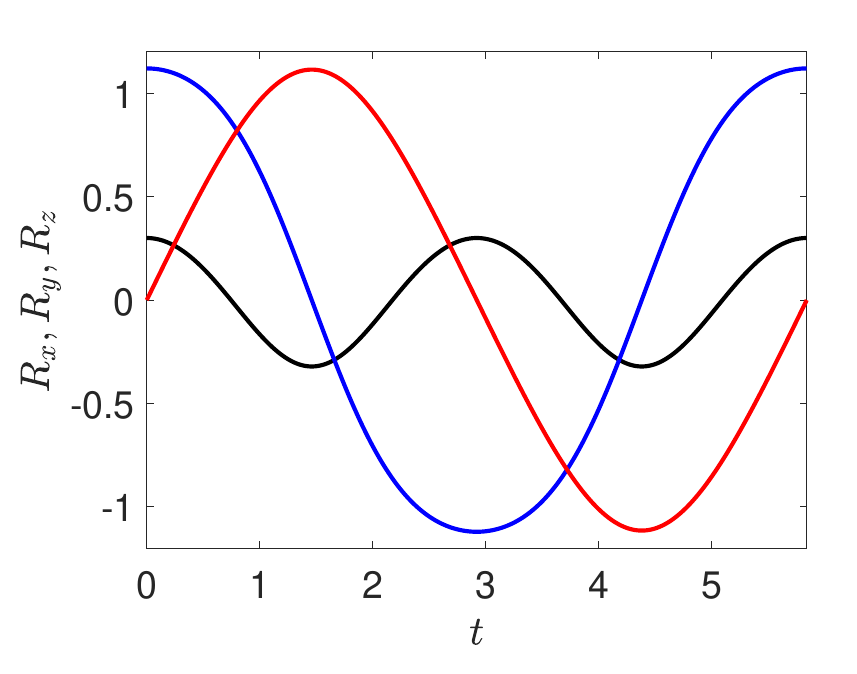}
\end{center}
\caption{Plot of the time evolution of $\vec{R}$ for $p_1=0.3002237=R_x(0)$, $p_2=-1.12045=-R_y(0)$ and $p_e=0=R_z(0)$ (see the text for details). The black, blue and red solid lines depict respectively $R_x$, $R_y$ and $R_z$. The time $t_f$ corresponds to one period of $R_z(t)$, i.e. $u_f=4K(m)$ and $t_f=u_f/A$.}
\label{fig1}
\end{figure}

We consider the singular situation with $\Delta=\Delta_s$ and $\Omega=\Omega_0$. The dynamics of $\vec{R}$ can be written as
\begin{equation}
    \dot{\vec{R}} = \Omega_0 \begin{pmatrix}
                0 & -\Omega_0R_z & 0 \\
                \Omega_0 R_z & 0 & -1 \\
                0 & 1 & 0
            \end{pmatrix}  \vec{R}.
\end{equation}
We consider the change of coordinates $\tilde{\vec{R}}=\Omega_0\vec{R}$ and $\tilde{t}=\Omega_0 t$ to obtain the following system of differential equations
\begin{equation}\label{eqR}
\begin{aligned}
\dot{R}_x&=-R_zR_y, \\
\dot{R}_y&=R_zR_x-R_z, \\
\dot{R}_z&=R_y,
\end{aligned}
\end{equation}
where we omit the tilde for simplicity. We denote by $\vec{R}^{(0)}$ the initial point of the dynamics. A careful inspection of Eq.~\eqref{eqR} shows that it has two constants of motion, $E_s$ and $R$ which can be expressed as
\begin{eqnarray*}
& & R^2={R_x}^2+{R_y}^2+{R_z}^2, \\
& & E_s=R_x+\frac{{R_z}^2}{2}.
\end{eqnarray*}
The values of such constants are fixed by the initial conditions on $\vec{R}$.
Starting from Eq.~\eqref{eqR}, we obtain that $R_z$ is solution of a  second order non-linear differential equation as
\begin{equation}
    \ddot{R}_z + \left(1 - E_s\right) R_z + \frac{1}{2} {R_z}^3 = 0.
\end{equation}
This can further be written has an effective Hamiltonian $E_z$ of a one-dimensional system as
\begin{equation}
    E_z = \frac{1}{2} {\dot{R}_z}^{~2} + \frac{1 - E_s}{2} {R_z}^2 + \frac{1}{8} {R_z}^4,
\end{equation}
which leads to
\begin{equation}
    {\dot{R}_z}^{~2} = -\frac{1}{4}\left[ {R_z}^4 + 4 \left(1 - E_s\right) {R_z}^2 - 8 E_z\right].
\end{equation}
Note that $E_z$ can be expressed in terms of the initial conditions $\vec{R}^{(0)}$. Since $E_z$ may be both positive or negative, the roots of the polynomial $P$ of order 2, $P(x)=x^2 + 4(1 - E_s) x - 8E_z$ can have the same or opposite signs (as its discriminant is always positive). In the numerical examples of Sec.~\ref{sec5} and \ref{sec6}, only the case $E_z>0$ (i.e. roots of opposite signs) allows to reach the target state and are thus relevant for our control problem. The solutions for which $E_z < 0$ are discussed in Appendix~\ref{appendix:Ez_neg}. We introduce the positive and negative roots of $P$ as $r^2$ and $-s^2$, with $r,s \in \R$, and
\begin{equation}
    \begin{aligned}
        r^2 &= -2 (1 - E_s) + 2 \sqrt{(1 - E_s)^2 + 2 E_z}, \\
        s^2 &= \phantom{-}2 (1 - E_s) + 2 \sqrt{(1 - E_s)^2 + 2 E_z}.
    \end{aligned}
\end{equation}
Note that $(1 - E_s)^2 + 2 E_z = (1 - R_x)^2 + {R_y}^2$. Introducing $\epsilon=\sgn(\dot{R}_z)=\sgn(R_y)$, we obtain
\begin{equation*}
    \int_{R_z^{(0)}}^{R_z(t)} \frac{du}{\sqrt{\left(r^2 - u^2\right)\left(s^2 + u^2 \right)}} = \epsilon\frac{t}{2},
\end{equation*}
and
\begin{equation}
    \label{eq:Rz sg general}
    R_z(t) = \frac{rs}{\sqrt{r^2 + s^2}} \mathrm{sd}\left(u = \frac{\sqrt{r^2 + s^2}}{2} t + u_0, m\right),
\end{equation}
where $\mathrm{sd} = \mathrm{sn} / \dn$ is a Jacobi elliptic function, and
\begin{equation}
    \begin{aligned}
        u_0 &= \mathrm{sd}^{-1} \left[\frac{\sqrt{r^2 + s^2}}{rs} R_z^{(0)}, m\right], \\
        m &= \frac{r^2}{r^2 + s^2}.
    \end{aligned}
\end{equation}
We introduce the parameter $A=\tfrac{\sqrt{r^2+s^2}}{2}$ connecting the coordinates $u$ and $t$. Note that the parameter $\epsilon$ is absorbed in a new definition of the sign of the product $rs$.
Using the relations $R_y = \dot{R}_z$ and $R_x = E_s - {R_z}^2 / 2$, we obtain for $\vec{R}$~\cite{abramovitz}
\begin{equation}
    \label{eq:R sg}
    \begin{aligned}
        R_x &= E_s + \frac{s^2}{2} - \frac{s^2}{2} \mathrm{nd}^2(u,m), \\
        R_y &= \frac{rs}{2} \frac{\cn(u,m)}{\dn^2(u,m)}, \\
        R_z &= \frac{rs}{\sqrt{r^2 + s^2}} \mathrm{sd}(u, m),
    \end{aligned}
\end{equation}
and for the NCR at time $t$
\begin{eqnarray}
& &    \mathcal{R}(t) = \left(E_s + \frac{s^2}{2}\right) t \\ & & - \frac{s^2}{\sqrt{r^2 + s^2}} \left[\Pi(\mu=m; \upsilon=u, m) - \Pi(\mu=m; \upsilon=u_0, m)\right] \nonumber
\end{eqnarray}
with $\Pi(\mu; u, m)$ the Jacobi incomplete elliptic integral of third kind. When $\mu = m$, this integral can be written as
\begin{equation}
    \Pi(m; u, m) = \frac{1}{1 - m} \left[E(u, m) - m \, \frac{\mathrm{sn}(u, m) \cn(u, m)}{\dn(u, m)}\right]
\end{equation}
with $E(u,m)$ the incomplete elliptic integral of the second kind~\cite{abramovitz}. Details of the analytical calculation can be found in Appendix~\ref{appendixellip}. An example for the time evolution of $\vec{R}$ is given in Fig.~\ref{fig1}.

    \subsection{The regular case}
The same derivation can be made in the regular case for which $\Delta=\Delta_0$ and $\Omega=\Omega_0$. The dynamics of $\vec{R}$ are given by
\begin{equation}
    \dot{\vec{R}} = \begin{pmatrix}
                0 & \Delta_0 & 0 \\
                -\Delta_0 & 0 & -\Omega_0 \\
                0 & \Omega_0 & 0
            \end{pmatrix}  \vec{R}
            \equiv M_0 \vec{R}.
\end{equation}
It can be shown that $R$ and $E_r=\Delta_0R_z-\Omega_0R_x$ are constants of motion. The propagator $e^{M_0 t}$ can be expressed as
$$
e^{M_0 t}=\1 + \left[1 - \cos\left(\omega t\right)\right] \frac{{M_0}^2}{\omega^2} + \sin\left(\omega t\right) \frac{M_0}{\omega},
$$
where $\1$ is the identity operator, and $\omega = \sqrt{{\Delta_0}^2 + {\Omega_0}^2}$. We deduce that the dynamic is $2\pi/\omega$-periodic. This gives $\vec{R}(t)=e^{M_0 t}\vec{R}(0)$
and
\begin{eqnarray}
& &    \mathcal{R}(t) = R_x^{(0)} t + \frac{\Delta_0}{\omega^3} \left[\sin(\omega t) - \omega t\right] \\
& & \times\left[\Delta_0 R_x^{(0)} + \Omega_0 R_z^{(0)}\right] + \frac{\Delta_0}{\omega^2} \left[\cos(\omega t) - 1\right] R_y^{(0)}\nonumber
\end{eqnarray}
for the NCR.
\section{Time-optimal state-to-state transfer}\label{sec5}
We begin by discussing the methodology used in the numerical examples. We consider an initial state $\ket{\psi_i}$ and a final target state $\ket{\psi_f}$, leading to initial and final states for $\vec{R}$, $\vec{R}(0) = \vec{R}_i$ and $\vec{R}(t_f) = \vec{R}_f$ using the fact that, for a given state $\ket{\psi} = (a, b)^\intercal$, $\vec{R}$ is given by
\begin{eqnarray}
\label{eq:R_qcq}
     \vec{R} &=& \begin{pmatrix}
            \Re\left(a^2 - b^2\right) \\
            -\Im\left(a^2 + b^2\right) \\
            -2 \Re (ab)
        \end{pmatrix} p_1 + \begin{pmatrix}
            -\Im\left(a^2 - b^2\right) \\
            -\Re\left(a^2 + b^2\right) \\
            2 \Im (ab)
        \end{pmatrix} p_2 \\
& &         + \begin{pmatrix}
        2 \Re (a^*b) \\
        2 \Im (a^*b) \\
        |a|^2 - |b|^2 \\
    \end{pmatrix} p_e.\nonumber
\end{eqnarray}
Note that $\vec{R}_i$ and $\vec{R}_f$ depend on the adjoint states $p_f=p_1+ip_2$ and $p_e$. Using the constants of motion, $R$, $E_s$ and $E_r$ we obtain several relations between such adjoint coordinates. We then use the conditions $\vec{R}(t_f)=\vec{R}_f$ and $\mathcal{R}(t_f)=0$ to find $t_f$ and the values of $(p_e,p_f)$. This calculation yields  candidates for robust optimality. Finally, we check whether such candidates are optimal solutions to the original control problem. We apply this procedure to several control problems.

\begin{figure}[htbp]
\begin{center}
\includegraphics[width=8cm]{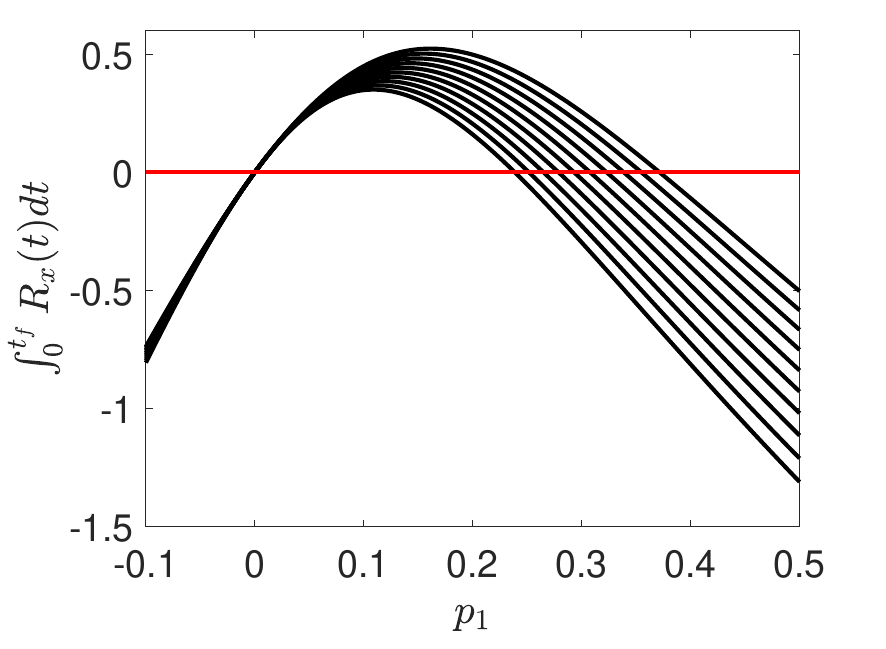}
\end{center}
\caption{Plot of $\mathcal{R}(t_f)=\int_0^{t_f}R_x(t)dt$ as a function of $p_1$ for different values of $\beta$ going from $-13.1\pi/12$ to $-12.9\pi/12$ (from left to right). The parameter $p_2$ is given by $p_2=-p_1\tfrac{1+\cos(2\beta)}{\sin(2\beta)}$. The horizontal red line indicates the axis of equation $\mathcal{R}(t_f)=0$.}
\label{fig2}
\end{figure}

\begin{figure}[htbp]
\begin{center}
\includegraphics[width=8cm]{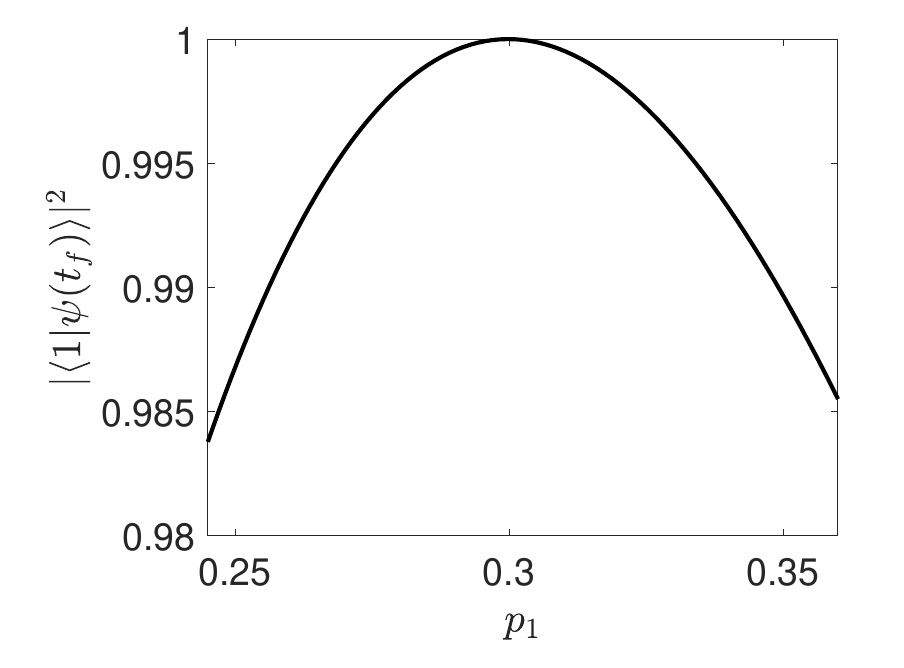}
\end{center}
\caption{Plot of the fidelity $|\langle \psi_f|\psi(t_f)\rangle|^2$ as a function of the set $(p_1,p_2)$ for which $\mathcal{R}(t_f)=0$. The optimal solution is obtained for $p_1=0.3002237$ and $p_2=-1.12045$.}
\label{fig3}
\end{figure}

We first consider the complete population transfer from $\ket{\psi_i} = \ket{0}$ to a state $\ket{\psi_f} = e^{i\beta}\ket{1}$. We emphasize that the final state is defined up to a global phase $\beta$ which is not fixed, as we are only interested in the population of the final state. Using Eq.~\eqref{eq:R_qcq} with $p_e = 0$, we derive the boundary conditions for $\vec{R}$
\begin{equation}
\label{eq:Ri and Rf total transfert}
    \vec{R}_i = \begin{pmatrix} p_1 \\ -p_2 \\ 0 \end{pmatrix},~
    \vec{R}_f = \begin{pmatrix}
            -\cos (2\beta) \, p_1 + \sin (2\beta) \, p_2 \\
            -\sin (2\beta) \, p_1 -\cos (2\beta) \, p_2 \\
            0
        \end{pmatrix}.
\end{equation}
In a first step, we assume that the solution is only singular on the interval $[0,t_f]$. Using the initial state of the dynamics, we determine the constants of motion as
\begin{equation}\label{constantmotion}
R=|p_f|,\ E_s=p_1,\ E_z=\frac{{p_2}^2}{2}.
\end{equation}
Note that, for the initial state studied here, $E_z > 0$ if $p_2 \neq 0$.
Furthermore, the condition $R_z(0)=0$ gives $u_0 = 0$ and
\begin{equation}
\label{eq:NCR sg total tranfer}
    \mathcal{R}(t_f) = \left(p_1 + \frac{s^2}{2}\right) t_f  - \frac{s^2}{\sqrt{r^2 + s^2}} \Pi(\mu=m; \upsilon=u_f, m).
\end{equation}
Using $R_z(t_f) = 0$ and the periodicity of the $\mathrm{sd}$ function, we obtain that the final value  $u_f$ must be of the form
\begin{equation}
    u_f = 2K(m) ~[2K(m)].
\end{equation}
All possible values of $u_f$ correspond to full periods for $R_x$, and either half periods or full periods for $R_y$ and $R_z$, respectively for $u_f \equiv 2K(m)$ and $u_f \equiv 4K(m)$. Each of these two possible values for $u_f$ verifies $R_x(t_f) = p_1 = R_x(0)$ using Eq.~\eqref{eq:R sg}, and adds the condition $p_2 \leq 0$ from $R_y(t_f)$ if we consider the case where $rs \geq 0$. Furthermore, these results from periodicity properties of the elliptic functions imply that the final point on the circle $R_z = 0$ of the $\vec{R}$-sphere is either $(R_x(0), -R_y(0))$ for $u_f \equiv 2K(m)$, or $(R_x(0), R_y(0))$ for $u_f \equiv 4K(m)$. This can be written as
\begin{equation}\label{eqp12}
    \begin{pmatrix} p_1 \\ \pm p_2 \end{pmatrix}
    =
    \begin{pmatrix}
            -\cos (2\beta) \, p_1 + \sin (2\beta) \, p_2 \\
            -\sin (2\beta) \, p_1 -\cos (2\beta) \, p_2
        \end{pmatrix},
\end{equation}
where the $\pm$ sign indicates the $2K(m)$ case up and the
$4K(m)$ case down. This leads to the following relations between $\beta$ and $p_f$
\begin{equation}
    \beta =
    \begin{cases}
        \pm\frac{\pi}{2} &u_f \equiv 2K(m) \\
        \frac{1}{2} \arctan\left(\frac{2 p_1 p_2}{{p_1}^2 + {p_2}^2}\right) &u_f \equiv 4K(m)
    \end{cases}.
\end{equation}
For a duration $2K(m)$, we verify numerically that for each value of $p_2$, a $p_1$ value can be found such that $\mathcal{R}(t_f)=0$. However, these candidates do not lead to a complete transfer in the original space. We thus focus only on the case $4K(m)$. Using Eq.~\eqref{eqp12}, we obtain that $p_2$ can be expressed as a function of $p_1$ and $\beta$. In this case, the corresponding trajectory reaches the target state at time $t_f$. The only remaining condition to satisfy is the NCR, $\mathcal{R}(t_f)=0$. Figure~\ref{fig2} represents the evolution of $\mathcal{R}(t_f)$ as a function of $p_1$ for different values of $\beta$. We observe that each curve has two intersection points with the horizontal axis $\mathcal{R}(t_f)=0$. The solution $p_1=0$ is trivial and we only consider the second solution. For each value of $\beta$, we obtain an optimal trajectory in the $\vec{R}$-space. Note that this set is parameterized by only one real parameter, $\beta$ or $p_1$. However, all these solutions do not perform the expected population transfer in the original space. Starting from the optimal control found in Fig.~\ref{fig2}, we plot in Fig.~\ref{fig3} the fidelity onto the excited state, $|\langle \psi_f|\psi(t_f)\rangle|^2$, as a function of $p_1$. Figure~\ref{fig3} shows that only one value of $p_1$ leads to a complete and robust population transfer. Numerically, we find $p_1=0.3002$ for $\beta=-13\pi/12$. Note that the same solution was obtained in~\cite{dridi2020} using the RIO approach.

\begin{figure}[htbp]
\begin{center}
\includegraphics[width=8cm]{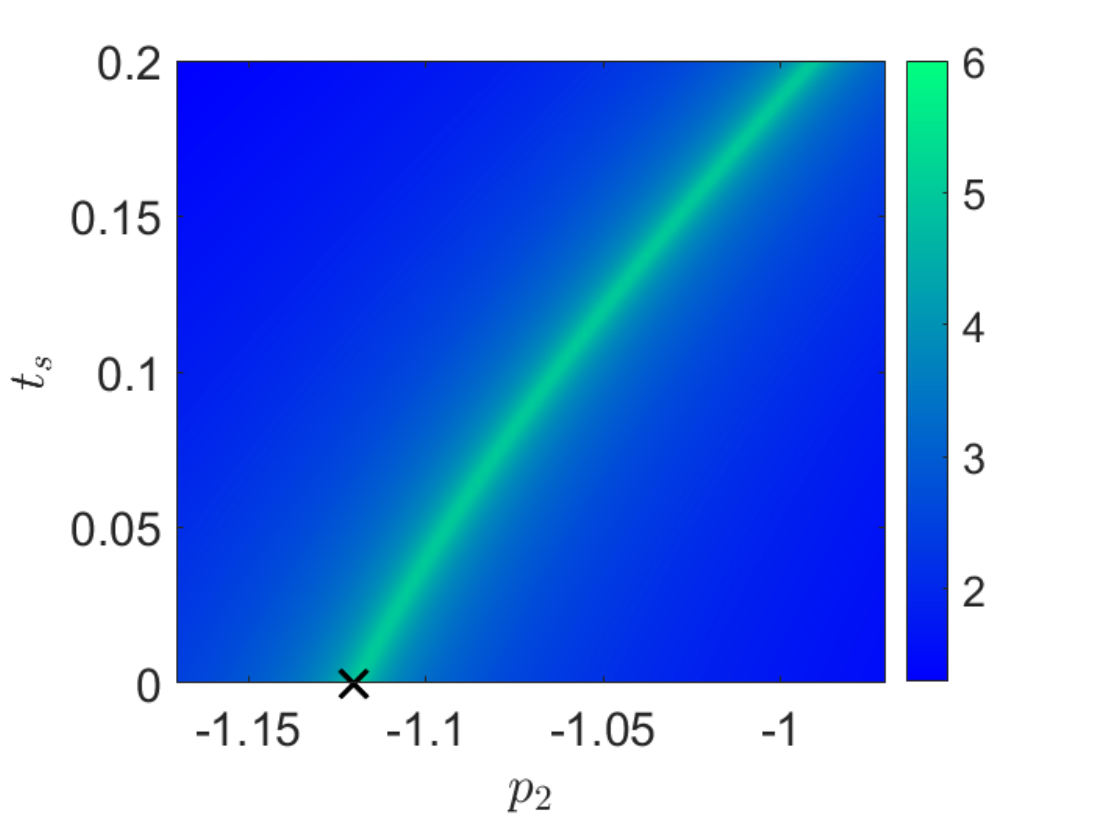}
\includegraphics[width=8cm]{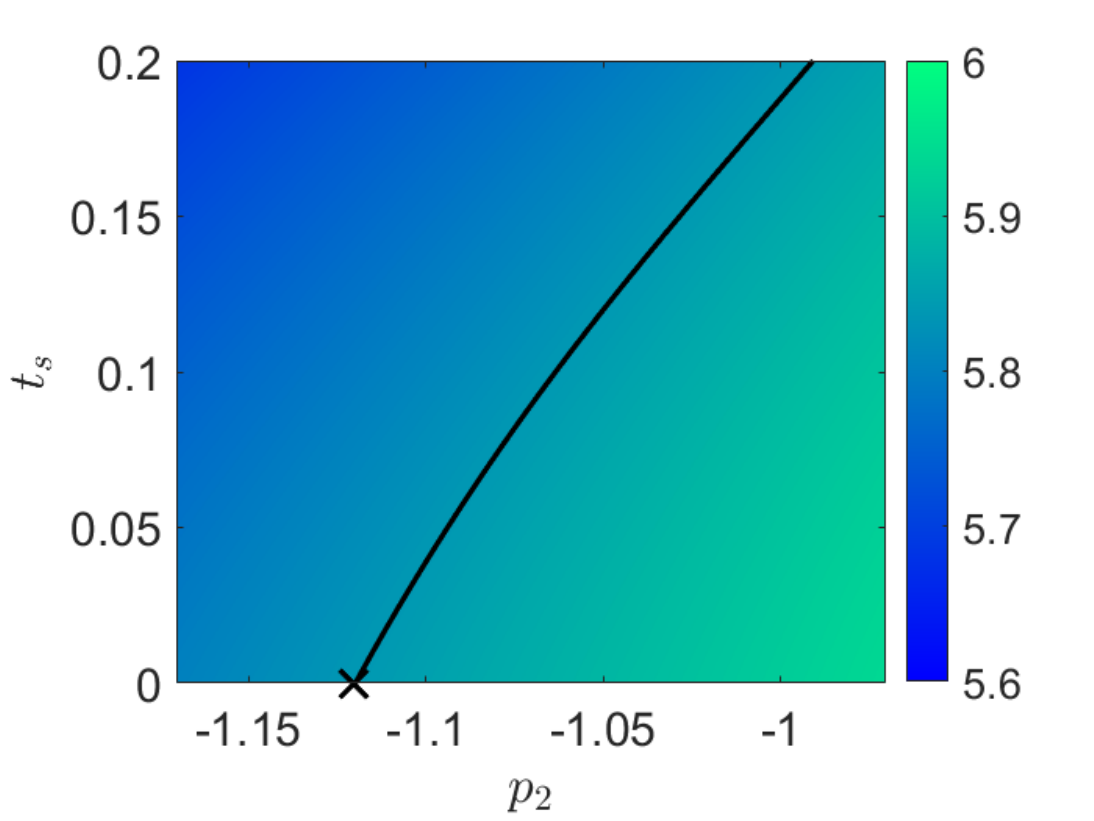}
\end{center}
\caption{(top) Plot of the fidelity $F_s$ as a function of $t_s$ and $p_2$. The regular control is set to $\Delta(t)=-\Delta_0=-1.5$. The adjoint state $p_1$ is found from the relation $\mathcal{R}(t_f)=0$. The black cross indicates the position of the time-optimal protocol with $p_1=0.3002237$, $p_2=-1.12045$ and $t_s=0$. (bottom) Control time $t_f$ as a function of $t_s$ and $p_2$. The solid black line represents the solutions for which $F_s>5$ and the black cross is the same as in the upper panel.}
\label{fig7}
\end{figure}

In a second step, we consider a solution that is a concatenation between a regular and a singular solution. Here, we assume that we start with a regular control from time $0$ to $t_s$ followed by a singular arc in the interval $[t_s,t_f]$. The constants of motion for the singular part are given by the final point of the regular arc. They become
\begin{eqnarray*}
        E_s &=& R_x(t_s)+\frac{R_z(t_s)^2}{2}, \\
        E_z &=& \frac{R_y(t_s)^2}{2}+\frac{1-E_s}{2}R_z(t_s)^2+\frac{R_z(t_s)^4}{8}.
\end{eqnarray*}
Since the final state does not change, the final value is still of the form $u_f=2K(m)~[2K(m)]$ to ensure that $R_z(t_f)=0$. We verify numerically that $2K(m)$ does not lead to a complete transfer and we focus on a control value of $4K(m)$. We have here an additional parameter to estimate corresponding to the switching time $t_s$. The boundary conditions on $\vec{R}$ are automatically satisfied when $u_f=4K(m)$. For each values of $p_2$ and $t_s$, we determine the value of $p_1$ for which $\mathcal{R}(t_f)=0$. We then obtain a set of solutions candidates to optimality. In a second step, we check numerically that such control laws lead to a complete population transfer in the original space. We introduce the fidelity $F_s=-\log(1-\mathcal{F}_s)=-\log(1-|b(t_f)|^2)$ as a measure of the distance to the target state. As displayed in Fig.~\ref{fig7}, a line of solutions is achieved as a function of $t_s$ and $p_2$. We select here the control protocols with $1-\mathcal{F}_s<10^{-5}$. A direct comparison of the correspond control times for such processes shows that the time-optimal control is the one for which $t_s=0$, i.e. with only a singular arc.

\begin{figure}[htbp]
\begin{center}
\includegraphics[width=8cm]{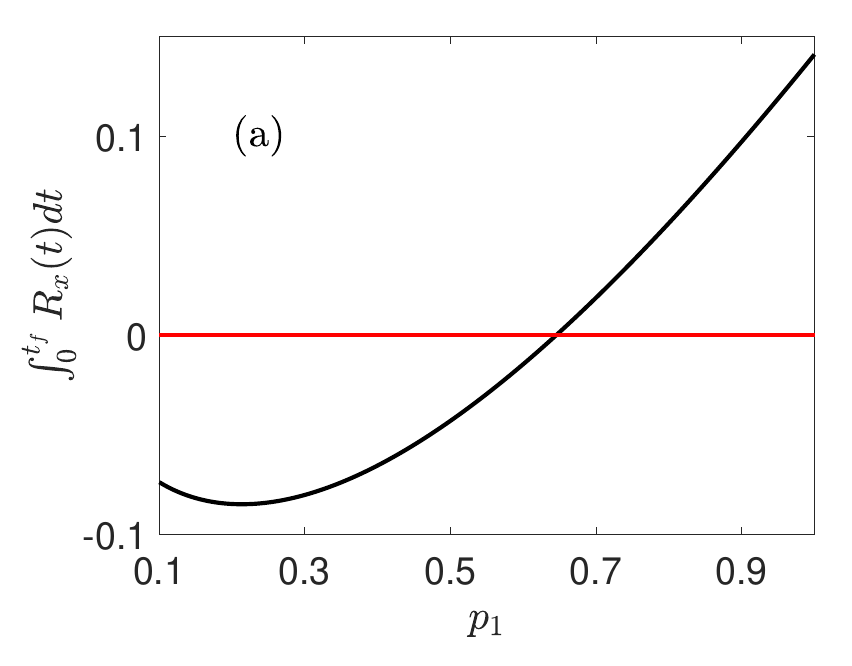}
\includegraphics[width=8cm]{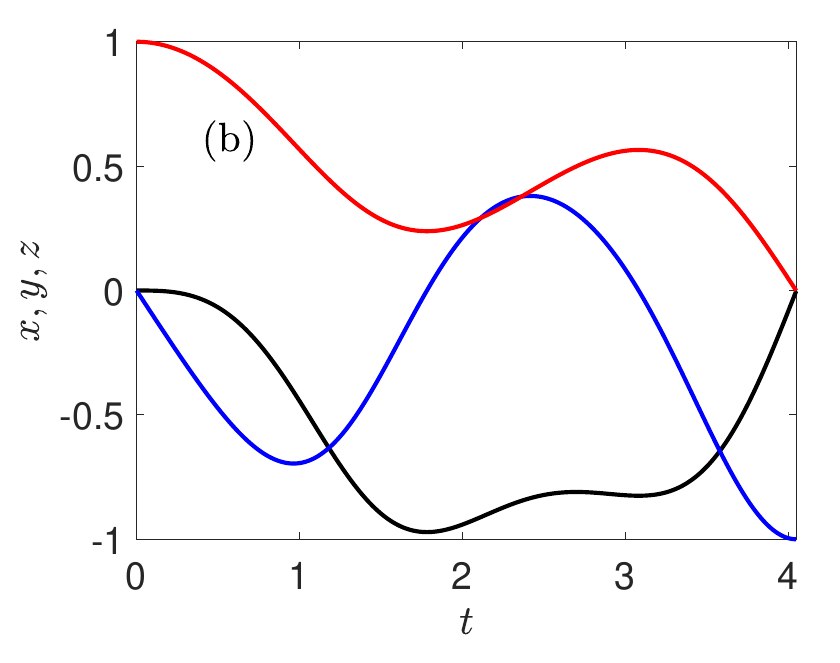}
\end{center}
\caption{Time-optimal solution for the transfer to a superposition of states. (a) Plot of the parameter $\mathcal{R}(t_f)$ as a function of $p_1$ for $\beta=0.74175\pi$. The equation of the horizontal red solid line is $\mathcal{R}(t_f)=0$. (b) Time evolution of the Bloch coordinates $x$, $y$ and $z$ for this optimal solution, with $x=\textrm{Tr}[|\psi_0\rangle\langle\psi_0|\sigma_{x}]$, $y=\textrm{Tr}[|\psi_0\rangle\langle\psi_0|\sigma_{y}]$ and $z=\textrm{Tr}[|\psi_0\rangle\langle\psi_0|\sigma_{z}]$. The black, blue and red solid lines display respectively the $x$, $y$ and $z$ coordinates.}
\label{fig4}
\end{figure}

As a second example, we now consider a half-transfer of population, along the $y$-axis of the Bloch sphere, from state $\ket{\psi_0} = \ket{0}$ to a state of the form $\ket{\psi_f} = [(1 + i) \ket{0} + (1 - i) \ket{1}] \, e^{i\beta} / 2$ where the global phase $\beta$ is not fixed. The initial state $\vec{R}_i$ is the same as in Eq.~\eqref{eq:Ri and Rf total transfert}, and the final state $\vec{R}_f$ is given by
\begin{equation}
    \vec{R}_f = \begin{pmatrix}
            -\sin (2\beta) \, p_1 - \cos (2\beta) \, p_2 \\
            0 \\
            -\cos (2\beta) \, p_1 + \sin (2\beta) \, p_2
        \end{pmatrix}.
\end{equation}
Note that this final state is the same as in Eq.~\eqref{eq:Ri and Rf total transfert}, up to a permutation of each term. Considering only a singular solution, the constants of motion are the same as in the first case (Eq.~\eqref{constantmotion}), $u_0 = 0$ and the NCR is also given by Eq.~\eqref{eq:NCR sg total tranfer}. Since $R_y(t_f) = 0$ and from the periodicity properties of the $\cn$ function, the final time must be of the form
\begin{equation}
    u_f = K(m) ~[2 K(m)].
\end{equation}
All possible values of $u_f$ correspond to half periods of $R_x$, and either quarter periods or three quarter periods of $R_y$ and $R_z$, respectively for $u_f \equiv K(m)$ and $u_f \equiv 3K(m)$.
Expressing $E_s$ at the initial and final times in terms of $p_1$ and $p_2$, we obtain
\begin{eqnarray*}
 E_s=p_1&=&-\sin(2\beta)p_1-\cos(2\beta)p_2\\
& & +\frac{1}{2}\left[-\cos(2\beta)p_1+\sin(2\beta)p_2\right]^2.
\end{eqnarray*}
For each value of $p_1$, we deduce that two values of $p_2^\pm$ may exist. They can be expressed as
\begin{eqnarray*}
p_2^\pm &=& \frac{1}{\sin^2(2\beta)}\left\{p_1\cos(2\beta)\sin(2\beta)+cos(2\beta) \phantom{\frac{}{}}\right. \\
& & \left.\pm \sqrt{\cos^2(2\beta)+2p_1\sin(2\beta)\left[1+\sin(2\beta)\right]}\right\}.
\end{eqnarray*}
The same analysis as in the complete population transfer can be performed. More precisely, for each value of $p_1$ and $\beta$, we obtain two possible values of $p_2$. First, we verify that the value $u_f=K(m)$ does not lead to solutions in the original space. In the second case, two equivalent solutions can be found: one for $p_2^-$, and the other for $p_2^+$. Numerically, we obtain $(p_1=0.64527,p_2^-=-1.69554,t_f=4.0479$) and $(p_1=0.64527,p_2^+=1.69554,t_f=4.0479$). The final values of the angle $\beta$ are respectively $0.74175\pi$ and $0.75825\pi$. The first set of solutions is represented in Fig.~\ref{fig4}. We use the Bloch coordinates $(x,y,z)$ to clearly show the transfer towards the $y$-axis on the Bloch sphere. Note that such solutions were also derived in~\cite{dridi2020} using the RIO approach.
\section{Quantum gates}\label{sec6}
In this section, we consider the realization of a quantum Not gate. The Hamiltonian $\hH$ of the system belongs to the Lie algebra $\mathfrak{su}(2)$, meaning that any evolution operator $\hU$ is an element of the Lie group $SU(2)$ and thus of the form
\begin{equation}
    \hU = \begin{pmatrix} a & -b^* \\ b & a^* \end{pmatrix}
    ,\
    |a|^2 + |b|^2 = 1.
\end{equation}
This operator is characterized by only two complex coefficients $a$ and $b$. The implementation of a one-qubit gate can thus be replaced by the study of the state  $\ket{\psi} = (a, b)^\intercal$.

\begin{figure}[htbp]
\begin{center}
\includegraphics[width=8cm]{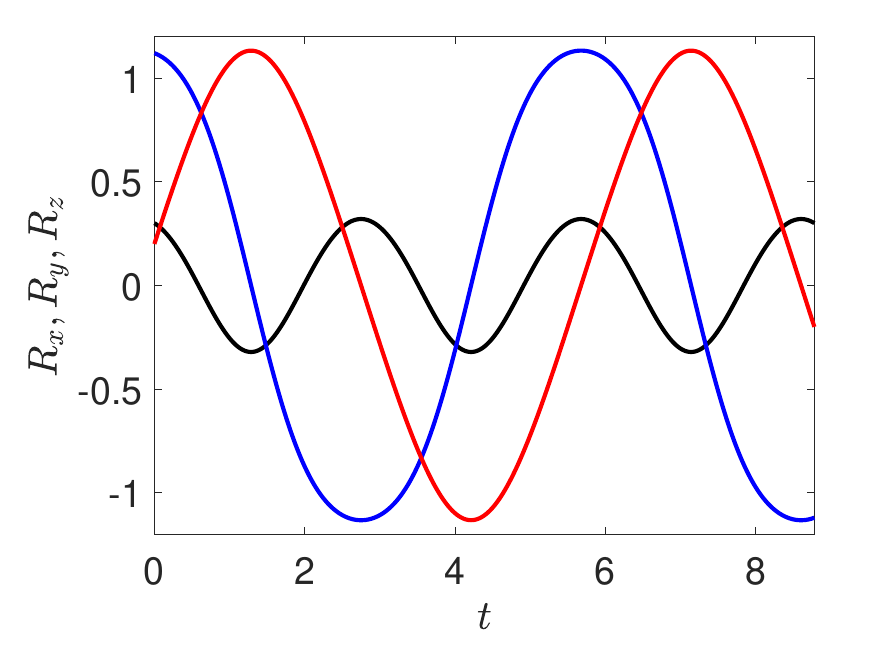}
\end{center}
\caption{Same as Fig.~\ref{fig1} but for $p_e=0.2$ and $t_f=6K(m)/A$.}
\label{fig5}
\end{figure}

\begin{figure}[htbp]
\begin{center}
\includegraphics[width=8cm]{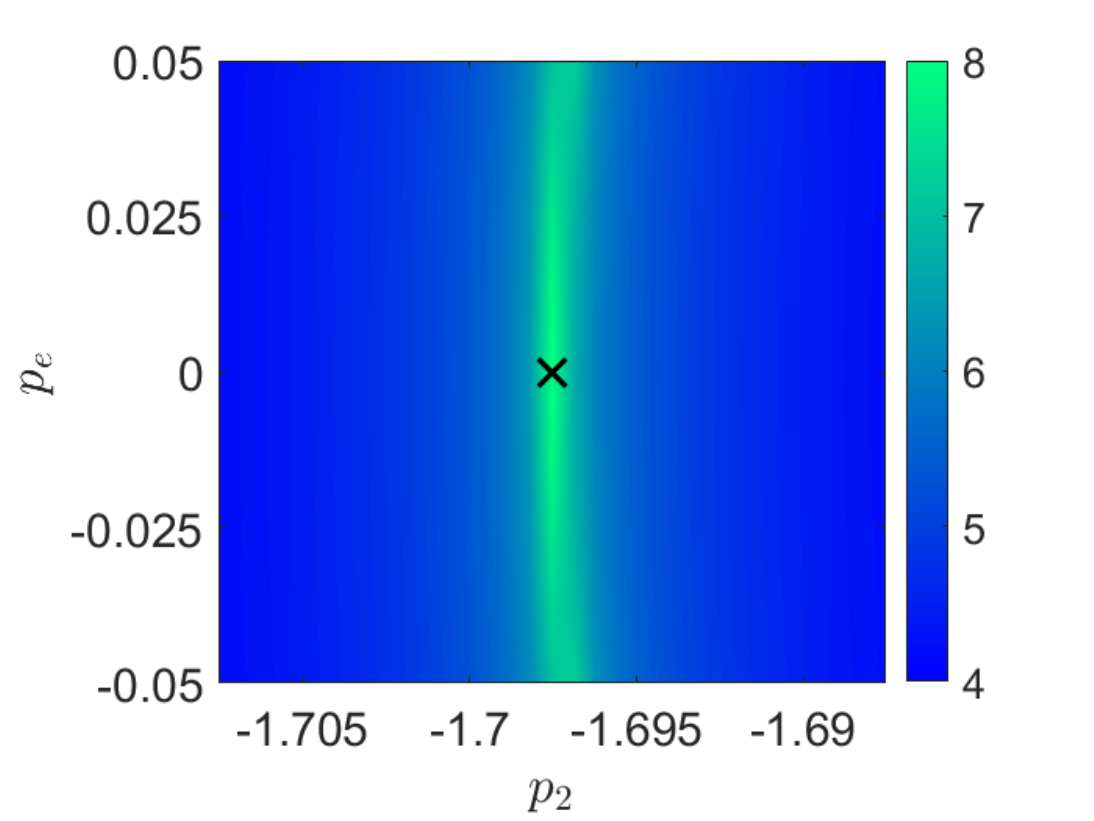}
\includegraphics[width=8cm]{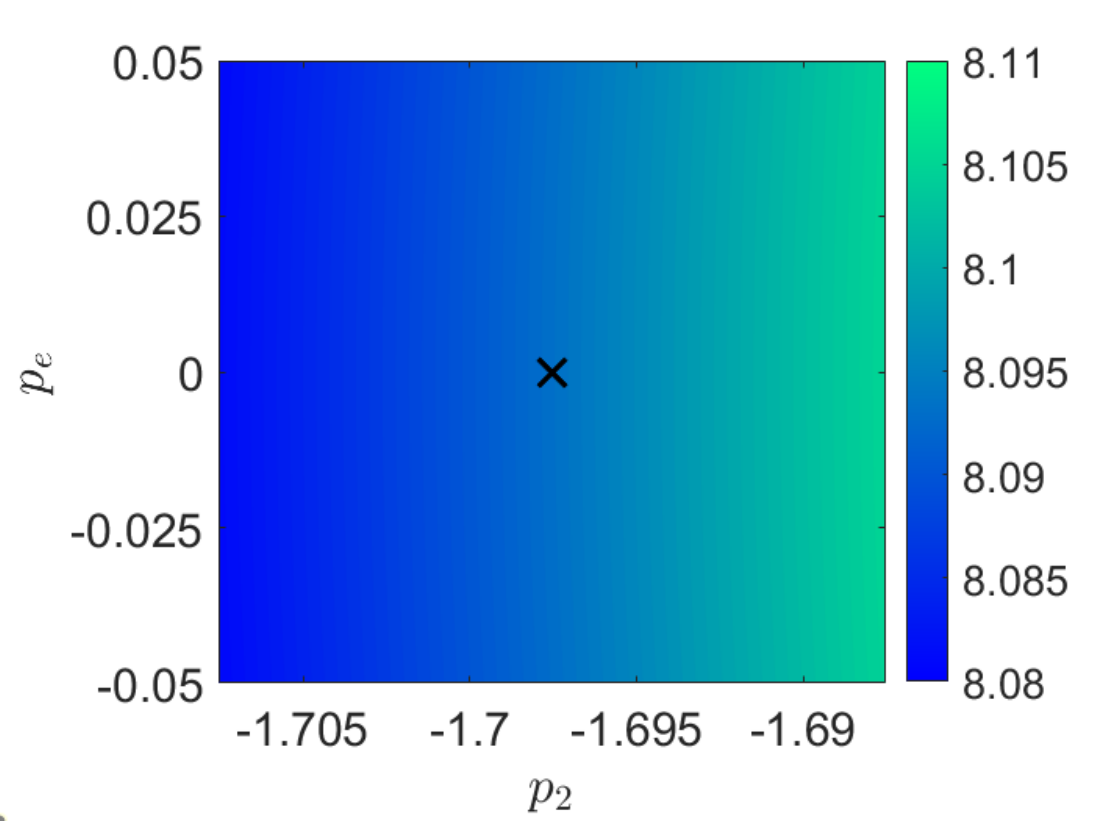}
\end{center}
\caption{(top) Plot of the figure of merit $F_g$ as a function of $p_2$ and $p_e$. The black cross indicates the position of the minimum value of $F_g$ for $p_2=-1.69754$ and $p_e=0$. (bottom) Plot of the corresponding minimum time $6K(m)/A$ with respect to $p_2$ and $p_e$. The minimum fidelity is reached for a time $t_f\simeq 8.093$}
\label{fig6}
\end{figure}

The usual Not gate, namely the $\Pauli{x}$ operator is of determinant -1 and cannot be reached by the dynamical system. Therefore, we consider the Not gate defined as $\hU_f = -i \Pauli{x}$. The initial state is the identity operator. The boundary conditions can be expressed as
\begin{equation}
    \begin{aligned}
        \hU_i &= \1, \\
        \hU_f &= -i\Pauli{x}
    \end{aligned}
    \implies
    \begin{aligned}
        \ket{\psi_i} &= \begin{pmatrix} 0 \\ 1 \end{pmatrix}, \\
        \ket{\psi_f} &= \begin{pmatrix} 0 \\ -i \end{pmatrix},
    \end{aligned}
\end{equation}
where the phase of the target state is fixed, as a different choice of global phase for $\ket{\psi_f}$ results in a different gate $\hU_f$. We then deduce the boundary conditions in $\vec{R}$ as
\begin{equation}
    \label{eq:Ri and Rf NOT gate}
    \vec{R}(0) = \begin{pmatrix} p_1 \\ -p_2 \\ p_e \end{pmatrix}
    ,\
    \vec{R}(t_f) = \begin{pmatrix} p_1 \\ p_2 \\ -p_e \end{pmatrix},
\end{equation}
and the constants of motion in the case of a singular control
\begin{equation}
    \begin{aligned}
        R^2 &= |p_f|^2 + {p_e}^2, \\
        E_s &= p_1 + \frac{{p_e}^2}{2}, \\
        E_z &= \frac{{p_2}^2}{2} + \frac{1 - p_1}{2} {p_e}^2 - \frac{{p_e}^4}{8}.
    \end{aligned}
\end{equation}
Using $R_z(0) = -R_z(t_f)$, we know from the periodicity and parity properties of the $\mathrm{sd}$ function that the final value must be of the form~\cite{abramovitz}
\begin{equation}\label{eqsol}
    u_f = A t_f + u_0 = \left\{ \begin{aligned}
            &2K(m) + u_0 & [4K(m)] \\
            &4K(m) - u_0 & [4K(m)]
        \end{aligned} \right.,
\end{equation}
with $A = \sqrt{r^2 + s^2} / 2$. We stress that the values of $u_f$ are respectively equal to $2K(m)~[4K(m)]$ and $(4K(m)-2u_0)~[4K(m)]$ in the upper and lower cases. When $u_0 = 0$ (i.e. $p_e=0$), the solutions are of the same form as in a state-to-state transfer with a singular control. Figure~\ref{fig5} illustrates the corresponding dynamics of $\vec{R}$ in the first case with $t_f=6K(m)/A$. A visual inspection shows that the boundary conditions~\eqref{eq:Ri and Rf NOT gate} are satisfied.
We first consider the upper solution in~\eqref{eqsol}. The goal is to determine the set of parameters $(p_1,p_2,p_e)$ to perform the unitary operation. Since $u_f - u_0 \equiv 2K(m)~[4K(m)]$ correspond to full periods of $R_x$, and half-periods of $R_y$ and $R_z$, the limit conditions are enforced by the periodicity and parity properties of the Jacobi elliptic functions, and the conditions on $\vec{R}$ are automatically satisfied. The adjoint state $p_1$ is then obtained as a function of $p_2$ and $p_e$ from the NCR, $\mathcal{R}(t_f)=0$. In the reduced space, all the corresponding solutions are valid and satisfy the final conditions. We then numerically check that such control protocols realize the expected operation in the original space. We introduce the figure of merit $F_g=-\log(1-\mathcal{F}_g)=-\log(1-\Im[b(t_f)])$ to measure the distance to the target. For a control value $u_f$ of $2K(m)$, we do not obtain a solution. Figure~\ref{fig6} represents the results for $t_f=6K(m)/A$. We find a very precise numerical solution for $p_2\simeq -1.69754$ and $p_e=0$. The corresponding value of $p_1$ is 0.6466 and $t_f\simeq 8.093$. Interestingly, we observe that this control process corresponds to a sequence of two state transfers leading to an equal superposition of the states $|1\rangle$ and $|2\rangle$. This general result is not surprising and was established in~\cite{luy2005}. A non-trivial point is that this solution corresponds to the time-optimal protocol, undoubtedly due to robustness constraints. This question requires a more general study that goes beyond the scope of this paper. Note that the same solution was derived in~\cite{dridi2020} with the RIO approach.

In the second case of Eq.~\eqref{eqsol}, the boundary conditions~\eqref{eq:Ri and Rf NOT gate} are satisfied if and only if $p_2=0$. Numerically, we do not find any values of $p_1$ and $p_e$ solutions of this problem in the $\vec{R}$-space.






\section{Conclusion}\label{sec8}
In this work, we apply the PMP to find robust time-optimal control protocols in a two-level quantum system. We consider both state-to-state transfers and the implementation of quantum gates. Introducing a reduced dynamical space, we show how the optimal equations given by the PMP can be integrated analytically. This derivation provides candidates to optimality for the control problem. The optimal solution is finally found by a numerical integration of the Schr\"odinger equation describing the system dynamics. In this paper, we focus on a specific class of robustness at order 2, that is the one with respect to Rabi frequency, but this approach can be generalized to other types of robustness and to higher orders following, for example, recent works in this direction~\cite{laforgue2022,dridi2024}. In such cases, it is not clear that the optimal solutions can be found analytically, but the reduced space method is expected to drastically simplify the numerical derivation of the control processes.

In the appendices~\ref{RIO} and~\ref{barnes}, we have also shown that two recent geometric optimization approaches used to solve this robustness problem are in fact specific formulations of the PMP and thus equivalent to this work. Such methods are interesting for highlighting the geometric nature of the control problem. However, they suffer from some limitations. The RIO approach only considers singular extremals and assumes additional smoothness conditions on the control. The use of angular coordinates in RIO can also lead to singularities that can be difficult to handle. The geometric optimization process proposed in~\cite{zeng2018,zeng2019} takes into account regular and singular extremals, but in a simplified situation that may be difficult to generalize.

Finally, we hope that our method can be used to solve more complex systems such as those with two or more coupled qubits. Similar approaches have been used, for example, in~\cite{pupillo2022} to generate gates in Rydberg atoms.

\noindent\textbf{Acknowledgment}\\
This research has been supported by the ANR project ``QuCoBEC'' ANR-22-CE47-0008-02 and by the ANR-DFG "CoRoMo" Projects No. 505622963/KO 2301/15-1 and No. ANR-22-CE92-0077-01.\\

\appendix

\section{Derivation of the optimal equations}\label{appder}
In this section, we show how the optimal equations of Sec.~\ref{sec3} can be derived.

We start with Eq.~\eqref{eqadj} which describes the dynamics of the adjoint state $|\chi\rangle$. We introduce the complex coordinates $a$ and $b$ so that the state $\ket{\psi_0}$ can be written as $\ket{\psi_0}= (a, b)^\intercal$ and an orthogonal state as $\ket{\psi_\perp} = (-b^*, a^*)^\intercal$. We have
\begin{equation*}
    \bra{\psi_\perp}\vPauli\ket{\psi_0} = \begin{pmatrix}
        a^2 - b^2 \\ ia^2 + ib^2 \\ 2ab
    \end{pmatrix}
    ,\
    \begin{aligned}
        \partial_{\ket{\psi_0}} \bra{\psi_\perp}\vPauli\ket{\psi_0} &= 2 \bra{\psi_\perp} \vPauli, \\
        \partial_{\ket{\psi_0}} \bra{\psi_0}\vPauli\ket{\psi_\perp} &= 0,
    \end{aligned}
\end{equation*}
and
\begin{equation*}
    \bra{\psi_0}\vPauli\ket{\psi_0} = \begin{pmatrix}
        2 \Re(a^* b) \\ 2i \Im(a^* b) \\ |a|^2 - |b|^2
    \end{pmatrix}
    ,\
    \partial_{\ket{\psi_0}} \bra{\psi_0}\vPauli\ket{\psi_0} = \bra{\psi_0}\vPauli.
\end{equation*}
We use here the fact that, for a scalar $G$, $\partial_{\ket{\psi_0}} G$ is a row vector such that
$$
\partial_{\ket{\psi_0}} G = \left(\partial_a G \ \ \partial_b G\right),
$$
where $a$, $b$ are independent from $a^*$, $b^*$. It is then straightforward to derive Eq.~\eqref{eqadj}.

Using the definitions of $\vec{I}$ and $\vec{R}$ in Eq.~\eqref{eqIR}, we show how to obtain Eq.~\eqref{eqderIR}. As an example, we consider $\vec{I}$. We have
$$
\dot{\vec{I}}=\Im\langle\dot{\chi}|\hat{\vec{\sigma}}|\psi_0\rangle + \Im\langle\chi|\hat{\vec{\sigma}}|\dot{\psi}_0\rangle.
$$
We deduce that
\begin{eqnarray*}
& & \dot{\vec{I}}=\Im(i\langle \chi|[H_0,\vec{\sigma}]|\psi_0\rangle)\\
& & -\Omega p_e \Im \langle\psi_0|\hat{\sigma}_x\hat{\vec{\sigma}}|\psi_0\rangle -\Omega \Im(p_f \langle\psi_\perp|\hat{\sigma}_x\hat{\vec{\sigma}}|\psi_0\rangle),
\end{eqnarray*}
which leads to Eq.~\eqref{eqderIR}.

\section{Integration of the dynamics of $\vec{R}$ for $E_z<0$}
\label{appendix:Ez_neg}
In this section, we present the calculation of the dynamics of $\vec{R}(t)$ for a singular control such that $E_z < 0$.

In this case, both roots of the polynomial $P(x) = x^2 + 4(1 - E_s) x - 8E_z$ are of the same sign. The sign of their sum is $\sgn(E_s - 1)$, thus it is also the sign of each root. We introduce the higher and lower roots of $P$ as $\sgn(E_s - 1) o^2$ and $\sgn(E_s - 1) q^2$, with $o, q \in \R$,  $o > q$, and
\begin{equation*}
    \begin{aligned}
        o^2 &= 2~ \sgn(E_s - 1) \left[(E_s - 1) + \sqrt{(E_s - 1)^2 + 2E_z}\right], \\
        q^2 &= 2~ \sgn(E_s - 1) \left[(E_s - 1) - \sqrt{(E_s - 1)^2 + 2E_z}\right].
    \end{aligned}
\end{equation*}
Since the sign of the roots is given by $\sgn(E_s - 1)$, the factorization of $P$ has a different form when $E_s < 1$ or $E_s > 1$.

We first consider the case $E_s < 1$, i.e. $\sgn(E_s - 1) = -1$. Both roots are negative, and the equation to solve is
\begin{equation*}
    {\dot{R}_y}^{~2} = - \frac{1}{4} \left({R_z}^2 + o^2\right) \left({R_z}^2 + q^2\right),
\end{equation*}
which gives that $\left({R_z}^2 + o^2\right) \left({R_z}^2 + q^2\right)$ must be negative, while being the sum and product of positive numbers. We deduce that there is no solution.

On the other hand, if $E_s > 1$, then $\sgn(E_s - 1) = 1$, and both roots are positive. The equation to solve becomes
\begin{equation*}
    \int_{R_z{(0)}}^{R_z(t)} \frac{du}{\sqrt{(o^2 - u^2)(u^2 - q^2)}} = \epsilon \frac{t}{2},
\end{equation*}
with $\epsilon = \sgn(\dot{R}_z) = \sgn(R_y)$, and we obtain for the control
\begin{equation*}
    R_z(t) = q \, \nd\left(u = \frac{o^2}{2} t + u_0, m\right),
\end{equation*}
with
\begin{equation}
\label{eq:Rz sg Ez<0}
     \begin{aligned}
        u_0 &= \nd^{-1}\left(\frac{R_z^{(0)}}{q}, m\right), \\
        m &= \frac{o^2 - q^2}{2}.
    \end{aligned}
\end{equation}
Note that, as in Eq.~\eqref{eq:Rz sg general}, the parameter $\epsilon$ is absorbed into a new definition of the sign of $q$. This results in $o > 0$.

In the case of a Not gate, we cannot have a change of sign for $R_z$, so the final condition of Eq.~\eqref{eq:Ri and Rf NOT gate} cannot be satisfied, and this gate cannot be constructed with $E_z < 0$.

\section{Analytical derivation of $\mathcal{R}(t)$}\label{appendixellip}
In this section, we describe how to find the analytical expression of $\mathcal{R}$ given in Sec.~\ref{sec4}.

We first recall the definition of the Jacobi incomplete elliptic integrals of first, second, and third order, respectively denoted $F(x| m)$, $E(x| m)$ and $\Pi(\mu; x| m)$, or as $F(u, m)$, $E(u, m)$ and $\Pi(\mu; u, m)$, with $u$ and $m$ being the arguments of the Jacobi functions, $\mu \in \R$~\cite{abramovitz}. They can be written as follows
\begin{align*}
    & \begin{aligned}
        F(x| m) &= \int_0^x \frac{dt}{\sqrt{(1 - t^2) (1 - mt^2)}}, \\
        E(x| m) &= \int_0^x \frac{\sqrt{1 - mt} \, dt}{\sqrt{1 - t^2}}, \\
        \Pi(\mu; x| m) &= \int_0^x \frac{dt}{(1 - \mu t^2) \sqrt{(1 - t^2) (1 - m t^2)}} ,
    \end{aligned}  \\
    & \begin{aligned}
        F(u, m) &= \int_0^u d\upsilon = u, \\
        E(u, m) &= \int_0^u \dn(\upsilon, m) \, d\upsilon, \\
        \Pi(\mu; u, m) &= \int_0^u \frac{d\upsilon}{1 - \mu \, \mathrm{sn}^2(\upsilon, m)}.
    \end{aligned}
\end{align*}
In the case where $x=1$, the integrals are called Jacobi complete elliptic integrals, and are denoted respectively, as $K(m)$, $E(m)$ and $\Pi(\mu; m)$.

We also point out that when $\mu = m$, $\Pi(\mu=m; u, m)$ can be expressed in terms of $E(u,m)$ as \cite{abramovitz}
\begin{equation}
    \Pi(m; u, m) = \frac{1}{1 - m} \left[E(u, m) - m \, \frac{\mathrm{sn}(u, m) \cn(u, m)}{\dn(u, m)}\right].
\end{equation}

The goal is then to integrate $R_x(t)$ in the singular case, given by Eq.~\eqref{eq:R sg}, to obtain an expression of the NCR $\mathcal{R}$ such as defined in Eq.~\eqref{eq:NCR}. We start by noting that $\dn^2(u,m) + m \, \mathrm{sn^2}(u, m) = 1$, such that $\int \nd^2(u,m) \, du = \int [1 - m \, \sn^2(u, m)]^{-1} \, du$. We obtain
\begin{equation*}
    \int_{u_0}^{u_f} \nd^2(u, m) \, du = \Pi(m; u_f, m) - \Pi(m; u_0, m).
\end{equation*}
Using $u = (\sqrt{r^2 + s^2} / 2) t + u_0$, we finally deduce that
\begin{equation*}
    \int_0^{t_f} \nd^2(u, m) \, dt = \frac{2}{\sqrt{r^2 + s^2}} \left[\Pi(m; u_f, m) - \Pi(m; u_0, m)\right],
\end{equation*}
which, for the NCR, leads to
\begin{align}
    &\mathcal{R}(t_f) = \left(E_s + \frac{s^2}{2}\right) t \\
        & - \frac{s^2}{\sqrt{r^2 + s^2}} \left[\Pi(m; u_f, m) - \Pi(m; u_0, m)\right]. \nonumber
\end{align}

\section{RIO and PMP}\label{RIO}
In this section, we study the connection between RIO and PMP. For the system investigated in this work, the RIO approach allows us to derive the singular solution of the PMP. We apply here the PMP by using angular coordinates as in~\cite{dridi2020}. This derivation highlights the link between the two methods. As an illustrative example, we consider a robust state-to-state transfer.

We parameterize the wave function in terms of the Euler angles $(\theta,\varphi,\gamma)$ with
$$
|\phi_0(t)\rangle =\begin{pmatrix}
\cos\frac{\theta}{2}e^{i\varphi/2} \\
\sin\frac{\theta}{2}e^{-i\varphi/2}
\end{pmatrix}
e^{-i\gamma/2}.
$$
The orthogonal state $|\phi_\perp\rangle$ can be written as
$$
|\phi_\perp(t)\rangle =\begin{pmatrix}
\sin\frac{\theta}{2}e^{i\varphi/2} \\
-\cos\frac{\theta}{2}e^{-i\varphi/2}
\end{pmatrix}
e^{i\gamma/2}.
$$
The dynamical equations are then given by
\begin{eqnarray*}
& &\dot{\theta}=\Omega \sin\varphi, \\
& & \dot{\varphi}=\Delta+\Omega\cos\varphi\cot\theta, \\
& & \dot{\gamma}=\Omega\frac{\cos\varphi}{\sin\theta},
\end{eqnarray*}
where $\Omega$ is assumed to be fixed and $\Delta(t)$ is the only control parameter. The goal of the control is a state-to-state transfer in the shortest possible time, e.g. from $\theta=0$ to $\theta=\pi$.

As shown in Sec.~\ref{sec2}, the control is robust at order 2 in $\alpha$ if $\left|\int_0^{t_f}f(t)dt\right|=0$ i.e. if $\Re\left[\int_0^{t_f}f(t)dt\right]=\Im\left[\int_0^{t_f}f(t)dt\right]=0$ where $f$ can be expressed as
$$
f(t)= \frac{\Omega}{2}e^{i\gamma}(i\sin\varphi-\cos\theta\cos\varphi).
$$
The two constraints can be expressed as
\begin{eqnarray*}
& & -\Re[F(t_f)]=\frac{\Omega}{2}\int_0^{t_f}[\sin\gamma\sin\varphi+\cos\gamma\cos\theta\cos\varphi]=0, \\
& & \Im[F(t_f)]=\frac{\Omega}{2}\int_0^{t_f}[\cos\gamma\sin\varphi-\sin\gamma\cos\theta\cos\varphi]=0.
\end{eqnarray*}
Such conditions can be rewritten using the angle $\gamma$ as the integration variable (here we assume that $\dot{\gamma}$ is a monotonic function). The function $f(t)$ can be expressed as
$$
f(t)=\frac{e^{i\gamma}}{2}(i\dot{\theta}-\frac{1}{2}\sin(2\theta)\dot{\gamma}).
$$
The first term of this function can be integrated by part. We have:
$$
\int_0^{t_f}f(t)dt=\frac{i\theta_f}{2}e^{i\gamma_f}-\frac{i\theta_i}{2}e^{i\gamma_i}-\frac{1}{4}\int_{\gamma_i}^{\gamma_f}d\gamma e^{i\gamma}(\sin(2\theta)-2\theta).
$$
The real and imaginary parts of the integral must be equal to 0. We deduce that
\begin{eqnarray*}
& & \frac{1}{4}\int_{\gamma_i}^{\gamma_f}d\gamma\cos\gamma (\sin(2\theta)-\theta)=\frac{1}{2}(\theta_i\sin\gamma_i-\theta_f\sin\gamma_f), \\
& & \frac{1}{4}\int_{\gamma_i}^{\gamma_f}d\gamma\sin\gamma (\sin(2\theta)-\theta)=-\frac{1}{2}(\theta_i\cos\gamma_i-\theta_f\cos\gamma_f).
\end{eqnarray*}
Parameterizing the trajectories by the angle $\gamma$, the dynamics are governed by the equations
\begin{eqnarray*}
& & \dot{\theta}=\tan\phi\sin\theta, \\
& & \dot{\phi}=\Delta\frac{\sin\theta}{\Omega\cos\phi}+\cos\theta,
\end{eqnarray*}
where the dot indicates the derivative with respect to $\gamma$.

In a time-optimal control process, the cost to minimize $\mathcal{C}$ is given by
$$
\mathcal{C}=\int_0^{t_f}dt=\frac{1}{\Omega}\int_{\gamma_i}^{\gamma_f}\frac{\sin\theta}{\cos\phi}d\gamma,
$$
where we assume that $\dot{\gamma}\geq 0$ and $\cos\phi\geq 0$, i.e. $-\frac{\pi}{2}\leq \phi\leq \frac{\pi}{2}$.
We introduce two new variables $\psi_1(\gamma)$ and $\psi_2(\gamma)$ as
\begin{eqnarray*}
& & \psi_1(\gamma)=-\frac{1}{4}\int_{\gamma_i}^\gamma d\gamma\sin\gamma(\sin(2\theta)-2\theta), \\
& & \psi_2(\gamma)=\frac{1}{4}\int_{\gamma_i}^\gamma d\gamma\cos\gamma(\sin(2\theta)-2\theta),
\end{eqnarray*}
with the final conditions
\begin{eqnarray*}
& & \psi_1(\gamma_f)=\frac{1}{2}(\theta_i\cos\gamma_i-\theta_f\cos\gamma_f), \\
& & \psi_2(\gamma_f)=\frac{1}{2}(\theta_i\sin\gamma_i-\theta_f\sin\gamma_f).
\end{eqnarray*}
We denote by $(p_1,p_2)$ the adjoint states. We emphasize that the robustness constraints are written in a slightly different form than in Sec.~\ref{sec2}. The Pontryagin Hamiltonian can be written as
\begin{eqnarray*}
H_P&=&p_\theta\tan\phi\sin\theta+p_\phi \left(\Delta\frac{\sin\theta}{\Omega\cos\phi}+\cos\theta\right)+p_0\frac{\sin\theta}{\cos\phi} \\
& &+p_1\dot{\psi}_1+p_2\dot{\psi}_2,
\end{eqnarray*}
where $p_0=-1$ in the normal case. We have:
\begin{eqnarray*}
H_P&=&p_\theta\tan\phi\sin\theta+p_\phi \left(\Delta\frac{\sin\theta}{\Omega\cos\phi}+\cos\theta\right)-\frac{\sin\theta}{\cos\phi}\\
&+&p_1\dot{\psi}_1+p_2\dot{\psi}_2.
\end{eqnarray*}
We deduce that the dynamical equations for the adjoint states are given by
\begin{eqnarray*}
& & \dot{p}_\theta = -p_\theta\tan\phi\cos\theta -p_\phi\left(\frac{\Delta}{\Omega}\frac{\cos\theta}{\cos\phi}-\sin\theta\right)\\
& & +\frac{\cos\theta}{\cos\phi}-\sin^2\theta(p_2\cos\gamma-p_1\sin\gamma),\\
& & \dot{p}_\phi = -\frac{p_\theta\sin\theta}{\cos^2\phi}-p_\phi\left(\frac{\Delta}{\Omega}\frac{\sin\theta\sin\phi}{\cos^2\phi}\right)\\
& & +\frac{\sin\theta\sin\phi}{\cos^2\phi}=
-\frac{\sin\theta}{\cos^2\phi}\left(p_\theta+p_\phi\sin\phi\frac{\Delta}{\Omega}-\sin\phi\right).
\end{eqnarray*}
We focus only on the singular extremals. We introduce the switching function $\Phi_\Delta$ defined as
$$
\Phi_\Delta=p_\phi\frac{\sin\theta}{\Omega\cos\phi}.
$$
The singular set satisfies $\Phi_\Delta=0$ over a non-zero time interval. A non-trivial solution is given by $p_\phi=0$. We deduce that $\dot{p}_\phi=0$, and
$$
-\dot{p}_\phi=\frac{\partial H_P}{\partial \phi}=0=\frac{p_\theta}{\cos^2\phi}\sin\theta-\sin\theta\frac{\sin\phi}{\cos^2\phi},
$$
with $p_\phi=0$. This leads to $p_\theta=\sin\phi$. The singular set is thus defined by the two conditions
$$
p_\phi=0,~p_\theta=\sin\phi.
$$
Note that this set corresponds in the space $(\cos\theta,\sin\phi,p_\theta,p_\phi)$ to a two-dimensional plane.  On this set, we have
\begin{eqnarray*}
& & -\dot{p}_\theta=-\cos\phi\cos\theta+\sin^2\theta(p_1\sin\gamma-p_2\cos\gamma), \\
& & \frac{d}{dt}\sin\phi=\cos\phi\dot{\phi}=\frac{\Delta\sin\theta}{\Omega}+\cos\theta\cos\phi.
\end{eqnarray*}
Using the fact that $d_t \sin\phi-\dot{p}_\theta=0$, we arrive at
$$
\Delta_s=\Omega\sin\theta(p_1\sin\gamma-p_2\cos\gamma).
$$
The singular control $\Delta_s$ is parameterized by the values of $(p_1,p_2)$. Plugging $\Delta_s$ into the dynamical equations, we get a singular dynamical system of the form
\begin{eqnarray*}
& & \dot{\theta}=\tan\phi\sin\theta, \\
& & \dot{\phi}=\frac{\sin^2\theta}{\cos\phi}(p_1\sin\gamma-p_2\cos\gamma)+\cos\theta.
\end{eqnarray*}
We deduce that
$$
\ddot{\theta}=\frac{\dot{\phi}}{\cos^2\phi}\sin\theta+\tan\phi\cos\theta\, \dot{\theta},
$$
which gives
$$
\ddot{\theta}=\frac{\sin^3\theta}{\cos^3\phi}(p_1\sin\gamma-p_2\cos\gamma)+\frac{\cos\theta\sin\theta}{\cos^2\phi}+\dot{\theta}^2\cot\theta,
$$
and
$$
\ddot{\theta}=\frac{\sin^3\theta}{\cos^3\phi}(p_1\sin\gamma-p_2\cos\gamma)+\cos\theta\sin\theta+2\dot{\theta}^2\cot\theta.
$$
Since
$$
\frac{1}{\cos^3\phi}=(1+\tan^2\phi)^{3/2}=\frac{1}{\sin^3\theta}(\sin^2\theta+\dot{\theta}^2)^{3/2},
$$
we get
\begin{equation}\label{eqRIO}
\ddot{\theta}=(p_1\sin\gamma-p_2\cos\gamma)(\sin^2\theta+\dot{\theta}^2)^{3/2}+\cos\theta\sin\theta+2\dot{\theta}^2\cot\theta.
\end{equation}
We stress that Eq.~\eqref{eqRIO} is the same equation as Eq.~(14) derived in Ref.~\cite{dridi2020}. In this case, this derivation explicitly shows that the RIO approach can be formulated from the PMP by considering only singular extremals.

\section{Geometric optimization and PMP}\label{barnes}
In this section, we consider the geometric optimization approach proposed in~\cite{zeng2018,zeng2019} from a PMP perspective. In order to highlight the connection between the two methods, we specifically study here the case of Ref.~\cite{zeng2018,zeng2019}.

We assume that the Hamiltonian $\hat{H}$ can be written as
$$
\hat{H}=-\frac{\Delta(t)}{2}\hat{\sigma}_z+\alpha\Omega \hat{\sigma}_x,
$$
where the control parameter is $\Delta(t)$ and $\Omega$ is set to 1. We introduce the phase $\phi(t)=\int_0^t\Delta(t')dt'$ and the following functions
$$
g_0(t)=1;~g_1(t)=\int_0^te^{-i\phi}\,dt';~g_2(t)=\int_0^te^{-i\phi}g_1^*\,dt';\dots
$$
As in Sec.~\ref{sec2}, we can expand the evolution operator in power series of $\alpha$, $\hat{U}(t,0)=\hat{U}_0(t,0)+\alpha\hat{U}_1(t,0)+\alpha^2\hat{U}(t,0)+O(\alpha^3)$. We have $\hat{U}_0(t,0)=\exp\left[i\phi(t)\frac{\Pauli{z}}{2}\right]$ and
$$
\hat{U}_1(t,0)=-i\begin{pmatrix}
0 & e^{i\varphi/2}g_1 \\
e^{-i\varphi/2}g_1^* & 0
\end{pmatrix},
$$
$$
\hat{U}_2(t,0)=-\begin{pmatrix}
e^{i\varphi/2}g_2 & 0 \\
0 & e^{i\varphi/2}g_2^*
\end{pmatrix}.
$$

Straightforward calculations lead to
\begin{eqnarray*}
& & a(t)=e^{i\phi /2}\left(g_0-g_2\alpha^2+\cdots \right), \\
& & b(t)=-ie^{-i\phi /2}(g_1^*\alpha+\cdots).
\end{eqnarray*}
The target unitary operator is a diagonal matrix given by $a_f=\exp(i\theta_f)$ at the final time $t_f$, and $b_f=0$. We have
\begin{eqnarray*}
& & \phi_f=2\theta_f+4k\pi, \\
& & g_1(t_f)=g_2(t_f)=\dots = 0,
\end{eqnarray*}
with the initial conditions $a(0)=1$ and $b(0)=\phi(0)=g_1(0)=g_2(0)=\dots =0$.
We introduce the interaction representation defined as $a=e^{i\phi/2}\tilde{a}$ and $b=e^{-i\phi/2}\tilde{b}$. We omit the tilde below and we expand $a$ and $b$ in power series of $\alpha$ as, e.g. $a=a_0+\alpha a_1+\alpha^2 a_2+\dots$. Using the following real coordinates $b_1=x_1+iy_1,~a_2=x_2+iy_2,\dots$, we arrive at
\begin{eqnarray*}
& & \dot{x}_1=\sin\phi, \\
& & \dot{y}_1=-\cos\phi, \\
& & \dot{x}_2=-\cos\phi y_1+\sin\phi x_1,\\
& & \dot{y}_2=\cos\phi x_1+\sin\phi y_1, \\
& & \cdots \\
& & \dot{\phi}=\Delta.
\end{eqnarray*}
We consider the robust optimal control at order 1 in minimum time. The differential system is restricted to the first two equations and to the last one, with the following initial and final conditions $x_1(0)=y_1(0)=x_1(t_f)=y_1(t_f)=0$ and $\phi(0)=0,~\phi(t_f)=\phi_f$.
The Pontryagin Hamiltonian can be expressed as
$$
H_P=p_1\sin\phi-q_1\cos\phi+p_\phi\Delta+p_0,
$$
where $(p_1,q_1,p_\phi)$ are the adjoint states of $(x_1,y_1,\phi)$ and $p_0=-1$. $p_1$ and $q_1$ are constants of motion and $p_\phi$ is solution of the equation
$$
\dot{p}_\phi=-p_1\cos\phi-q_1\sin\phi.
$$
Since the final time is free, $H_p=0$. The switching function $\Phi_\Delta$ is $\Phi_\Delta=p_\phi$.

When $\Phi\neq 0$, we have a regular extremal and $\Delta=\varepsilon \Delta_0$ with $\varepsilon=\pm 1$ and $\phi=\varepsilon \Delta_0 t+\phi_0$. We deduce that
\begin{eqnarray*}
& & x_1(t)=-\frac{1}{\varepsilon \Delta_0}\cos(\varepsilon \Delta_0 t+\phi_0)+x_1^{(0)}, \\
& & y_1(t)=-\frac{1}{\varepsilon \Delta_0}\sin(\varepsilon \Delta_0 t+\phi_0)+y_1^{(0)}.
\end{eqnarray*}
The corresponding trajectories are circles of radius $1/\Delta_0$.

In the singular case, the switching function is zero on a non-zero time interval. In this interval, $\dot{p}_\phi=0$ and we get
$$
\dot{p}_\phi=-p_1\cos\phi-q_1\sin\phi=0,
$$
which leads to $\tan\phi=-\tfrac{p_1}{q_1}$, $\phi(t)$ is a constant and $\Delta_s=0$. On a singular extremal, we know that
\begin{eqnarray*}
& & p_1\sin\phi-q_1\cos\phi =1, \\
& & -p_1\cos\phi-q_1\sin\phi=0,
\end{eqnarray*}
which leads to $q_1=-\cos\phi$ and $p_1=\sin\phi$. Note that singular extremals can be used only if ${p_1}^2+{q_1}^2=1$. We denote by $\phi_s$ the constant value of $\phi$ on a singular extremal. The corresponding trajectories are straight lines in the $(x_1,y_1)$-space with
\begin{eqnarray*}
& & x_1(t)=\sin(\phi_s)\, t+x_1^{(0)}, \\
& & y_1(t)=-\cos(\phi_s)\, t+y_1^{(0)}.
\end{eqnarray*}
The optimal solution can be the concatenation of bang and singular arcs.

We assume that $\Delta(0)=\Delta_0$. We have $\phi(t)=\Delta_0t$ and
\begin{eqnarray*}
& & x_1(t)=-\frac{1}{\Delta_0}\cos(\Delta_0t)+\frac{1}{\Delta_0}, \\
& & y_1(t)=-\frac{1}{\Delta_0}\sin(\Delta_0t).
\end{eqnarray*}
We obtain $(x_1-\tfrac{1}{\Delta_0})^2+{y_1}^2=\tfrac{1}{{\Delta_0}^2}$. The initial point $(x_1,y_1)=(0,0)$ belongs to a circle of radius $\tfrac{1}{\Delta_0}$ and of center $(\tfrac{1}{\Delta_0},0)$. For the adjoint state $p_\phi(t)$, we get
$$
p_\phi(t)=p_\phi^{(0)}-\frac{p_1}{\Delta_0}\sin(\Delta_0 t)+\frac{q_1}{\Delta_0}\cos(\Delta_0 t).
$$
We set
\begin{eqnarray*}
& & \frac{q_1}{\sqrt{{p_1}^2+{q_1}^2}}=-\cos\theta, \\
& & \frac{p_1}{\sqrt{{p_1}^2+{q_1}^2}}=\sin\theta,
\end{eqnarray*}
and we arrive at
$$
p_\phi(t)=p_\phi^{(0)}-\frac{\sqrt{{p_1}^2+{q_1}^2}}{\Delta_0}\cos(\Delta_0t-\theta).
$$
The singular set is reached when $p_\phi(t)=0$ and if ${p_1}^2 + {q_1}^2 = 1$, which gives the condition

$$
\cos(\Delta_0 t-\theta)= p_\phi^{(0)}\Delta_0 =\cos\eta,
$$
that is valid if $|p_\phi^{(0)}|\leq \tfrac{1}{\Delta_0}$. A switching occurs if either
\begin{eqnarray*}
\Delta_0 t = \theta+\eta~ [2\pi], \\
\Delta_0 t = \theta-\eta~ [2\pi].
\end{eqnarray*}
We recover here exactly the solutions derived in~\cite{zeng2018,zeng2019}.

\end{document}